\documentclass[twocolumn,floatfix]{revtex4-2}
\pdfoutput=1
\usepackage{graphicx}
\usepackage{dcolumn}
\usepackage{longtable}
\usepackage{amsmath}
\usepackage{amssymb}
\usepackage{color}
\usepackage{isotope}

\begin{document}
\title{Prolate-oblate shape transitions and O(6) symmetry in even-even nuclei: A theoretical overview}

\author
{Dennis Bonatsos$^1$, Andriana Martinou$^1$,  S.K. Peroulis$^1$, T.J. Mertzimekis$^2$, and N. Minkov$^3$ }

\affiliation
{$^1$Institute of Nuclear and Particle Physics, National Centre for Scientific Research ``Demokritos'', GR-15310 Aghia Paraskevi, Attiki, Greece}

\affiliation
{$^2$  Department of Physics, National and Kapodistrian University of Athens, Zografou Campus, GR-15784 Athens, Greece}

\affiliation
{$^3$Institute of Nuclear Research and Nuclear Energy, Bulgarian Academy of Sciences, 72 Tzarigrad Road, 1784 Sofia, Bulgaria}

\begin{abstract}

Prolate to oblate shape transitions have been predicted in an analytic way in the framework of the Interacting Boson Model (IBM), determining O(6) as the symmetry at the critical point. Parameter-independent predictions for prolate to oblate transitions in various regions on the nuclear chart have been made in the framework of the proxy-SU(3) and pseudo-SU(3) symmetries, corroborated by recent non-relativistic and relativistic mean field calculations along series of nuclear isotopes, with parameters fixed throughout, as well as by shell model calculations taking advantage of the quasi-SU(3) symmetry. Experimental evidence for regions of prolate to oblate shape transitions is in agreement with regions in which nuclei bearing the O(6) dynamical symmetry of the IBM have been identified, lying below major shell closures. In addition, gradual oblate to prolate transitions are seen when crossing major nuclear shell closures, in analogy to experimental observations in alkali clusters.  

 \end{abstract}

\maketitle

{Corresponding author: Dennis Bonatsos. E-mail: bonat@inp.demokritos.gr} 

\section{Introduction}   

Atomic nuclei are characterized by a large variety of shapes: spherical \cite{ScharffGoldhaber1955}, axially symmetric ellipsoidal \cite{Rainwater1950} (having as special cases the superdeformed \cite{Twin1986,Nolan1988} and hyperdeformed \cite{LaFosse1995,LaFosse1996} nuclei, with axes ratios 2:1 and 3:1 respectively), triaxial \cite{MeyerterVehn1974,MeyerterVehn1975}, unstable towards triaxial deformation (called $\gamma$-unstable) \cite{Wilets1956}, octupole (pear-like) \cite{Butler1996,Butler2016}, hexadecapole \cite{Burke1994,Garrett2005}. Axially symmetric  ellipsoidal nuclei are divided into prolate deformed (rugby-ball like, with two short axes and a long one) and oblate deformed (pancake like, with two long axes and a short one) \cite{Nilsson1955,Ragnarsson1978,Nilsson1995}. Most even-even nuclei possess prolate ground states, with oblate ground states seen in few of them. This prolate over oblate dominance has been a challenge for theoretical approaches for many years \cite{Hamamoto2009,Hamamoto2012}. 

Transitions from one shape to another along series of isotopes of even-even nuclei have been studied for a long time \cite{Scholten1978,Rowe2004b,Turner2005,Rosensteel2005}. When the nuclear shape changes abruptly by the addition of a neutron pair, the term shape/phase transition (SPT) has been used \cite{Iachello2006}, with the neutron number playing the role of the control parameter of the transition \cite{Casten1999}. A first-order (abrupt) SPT  has been observed from spherical to prolate deformed nuclei \cite{Feng1981,Iachello2001}, while a second-order (less abrupt) SPT has been seen between spherical and $\gamma$-unstable nuclei \cite{Feng1981,Iachello2000}. The question of the possible existence of a SPT from prolate to oblate shapes, as well as of the order of such a SPT, have also been posed since several years \cite{LopezMoreno1996,Bonatsos2004}.

The theoretical framework for the microscopic study of atomic nuclei in terms of their constituent protons and neutrons has been set by the introduction of the nuclear shell model in 1949 \cite{Mayer1948,Mayer1949,Haxel1949,Mayer1955}, while the framework for the macroscopic study of the variety of nuclear shapes has been formed through the introduction of the collective model of Bohr and Mottelson in 1952 \cite{Bohr1952,Bohr1953,Bohr1998a,Bohr1998b}. 

An alternative microscopic approach is provided by the nuclear mean field methods \cite{Bender2003}, both non-relativistic \cite{Delaroche2010,Erler2011} and relativistic
\cite{Ring1997,Lalazissis1997,Lalazissis2005,Niksic2014}, the latter being based on density functional theory approaches \cite{Dobaczewski2011,Dobaczewski2012}, first developed for the study of many-electron systems \cite{Hohenberg1964,Kohn1965}. In these cases the nuclear mean field is obtained through fitting to the nuclear structure data, in contrast to the shell model, in which a specific single-particle potential is assumed, followed by configuration mixing within the shell model space taken into account. 

Another alternative path is provided by taking advantage of group theory, since the importance of symmetries in nuclear structure has been recognized by Wigner already in 1937 \cite{Wigner1937}. A bridge between the shell model and the collective model has been provided in 1958 by the SU(3) model of Elliott \cite{Elliott1958a,Elliott1958b,Elliott1963,Elliott1968}, on which the pseudo-SU(3) \cite{Arima1969,Hecht1969,RatnaRaju1973}, quasi-SU(3) \cite{Zuker1995,Zuker2015}, and proxy-SU(3) \cite{Bonatsos2017a,Bonatsos2017b,Bonatsos2023a} models have been later built. The microscopic symplectic model \cite{Rosensteel1980,Rowe1985}, as well as its recent extension, the symmetry-adapted no-core shell model \cite{Dytrych2008b,Launey2016,Launey2021}, also make wide use of the SU(3) symmetry techniques \cite{Kota2020}.

In addition, a phenomenological algebraic approach to nuclear structure has been introduced in 1975 by Arima and Iachello \cite{Arima1975}, consisting of the Interacting Boson Model and its several extensions \cite{Iachello1987,Iachello1991}. Within this model, axially deformed nuclei are described by the SU(3) dynamical symmetry \cite{Arima1978b}, while spherical and $\gamma$-unstable nuclei correspond to the U(5) \cite{Arima1976} and O(6) \cite{Arima1979} dynamical symmetries respectively. The classical limit of IBM \cite{Ginocchio1980a,Ginocchio1980b,Dieperink1980}, obtained at large boson numbers, provides a bridge with the collective model of Bohr and Mottelson. As we shall see in the present review, several interconnections exist between the O(6) symmetry and the prolate to oblate SPT.        

The scope of the present review is to gather together theoretical predictions for prolate to oblate transitions made in the framework of several different approaches, namely algebraic models using bosons (sec. \ref{algebraic}), special solutions of the Bohr collective model (sec.\ref{Bohr}), non-relativistic and relativistic mean field (RMF) models (sec. \ref{meanfield}), as well as the nuclear shell model and its pseudo-SU(3), quasi-SU(3) and proxy-SU(3) approximations (sec. \ref{shellmodel}). The specific predictions of these approaches in various regions of the nuclear chart are arranged by series of isotopes (sec. \ref{po}) and compared to regions in which manifestations of the O(6) dynamical symmetry (DS) of IBM appear (sec. \ref{O(6)}), with good agreement found between the two sets. This interrelation is used in order to clarify the nature of the O(6) DS, as well as the nature of the oblate to prolate transition observed when crossing major shell closures, in analogy to alkali clusters (secs. \ref{O6U5} to \ref{Qoper2}). For a more detailed description of the theoretical approaches used, the reader is referred to section 2 of the recent review article \cite{Bonatsos2023c}. 
The acronyms used in the text are listed in an Appendix. 

\section{Algebraic models using bosons} \label{algebraic}

The Interacting Boson Model (IBM) \cite{Arima1975,Iachello1987,Iachello1991}, describing the collective properties of nuclei in terms of $s$ ($L=0$) and $d$ ($L=2$) bosons,  has been for decades the most popular algebraic nuclear model. Having an overall U(6) symmetry, it possesses three dynamical symmetries (DSs), U(5) \cite{Arima1976}, SU(3) \cite{Arima1978b}, and O(6) \cite{Arima1979}, corresponding to vibrational, axially deformed, and $\gamma$-unstable (soft to triaxial deformation) nuclei respectively. 

\begin{figure*}[htb] 

{\includegraphics[width=75mm]{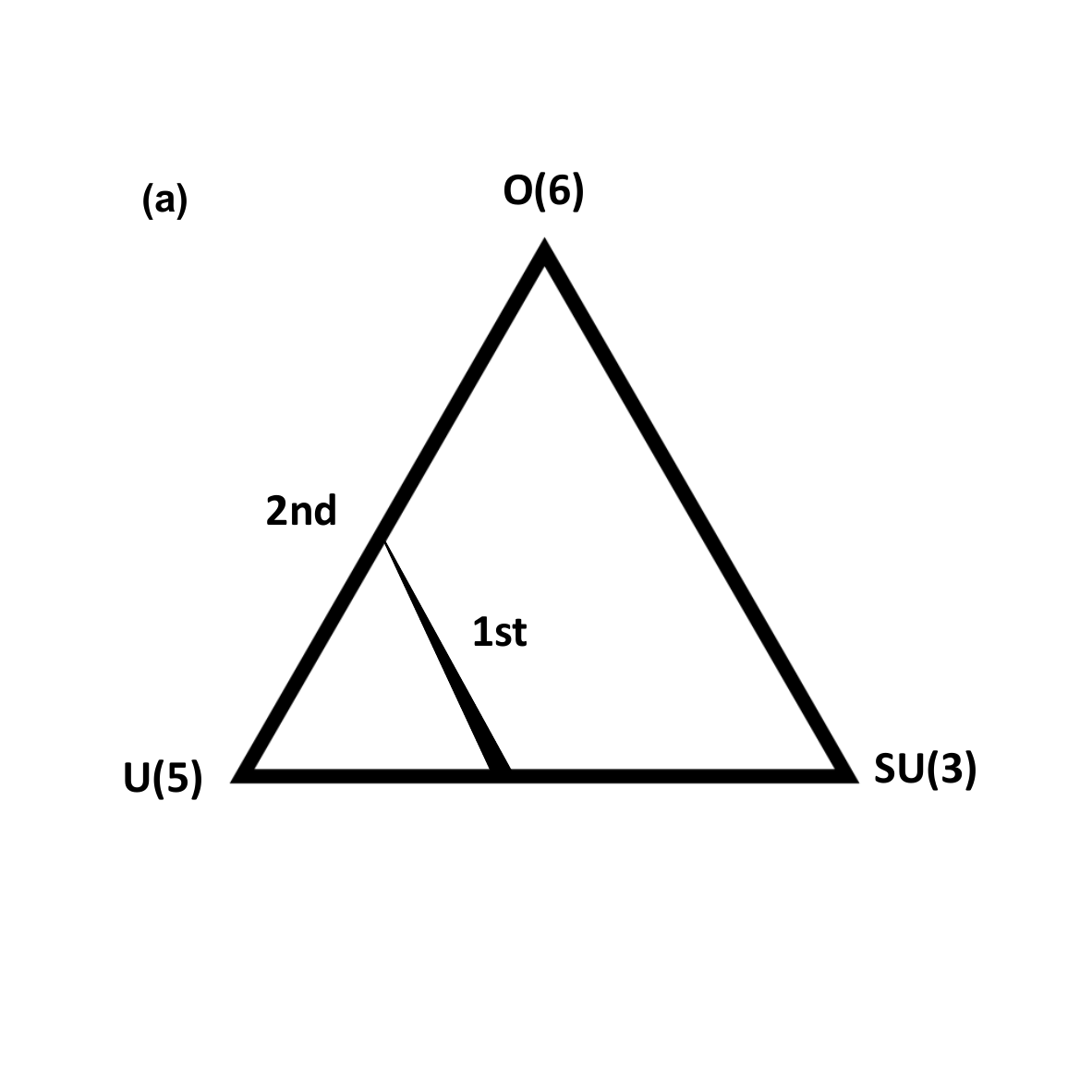}
\includegraphics[width=75mm]{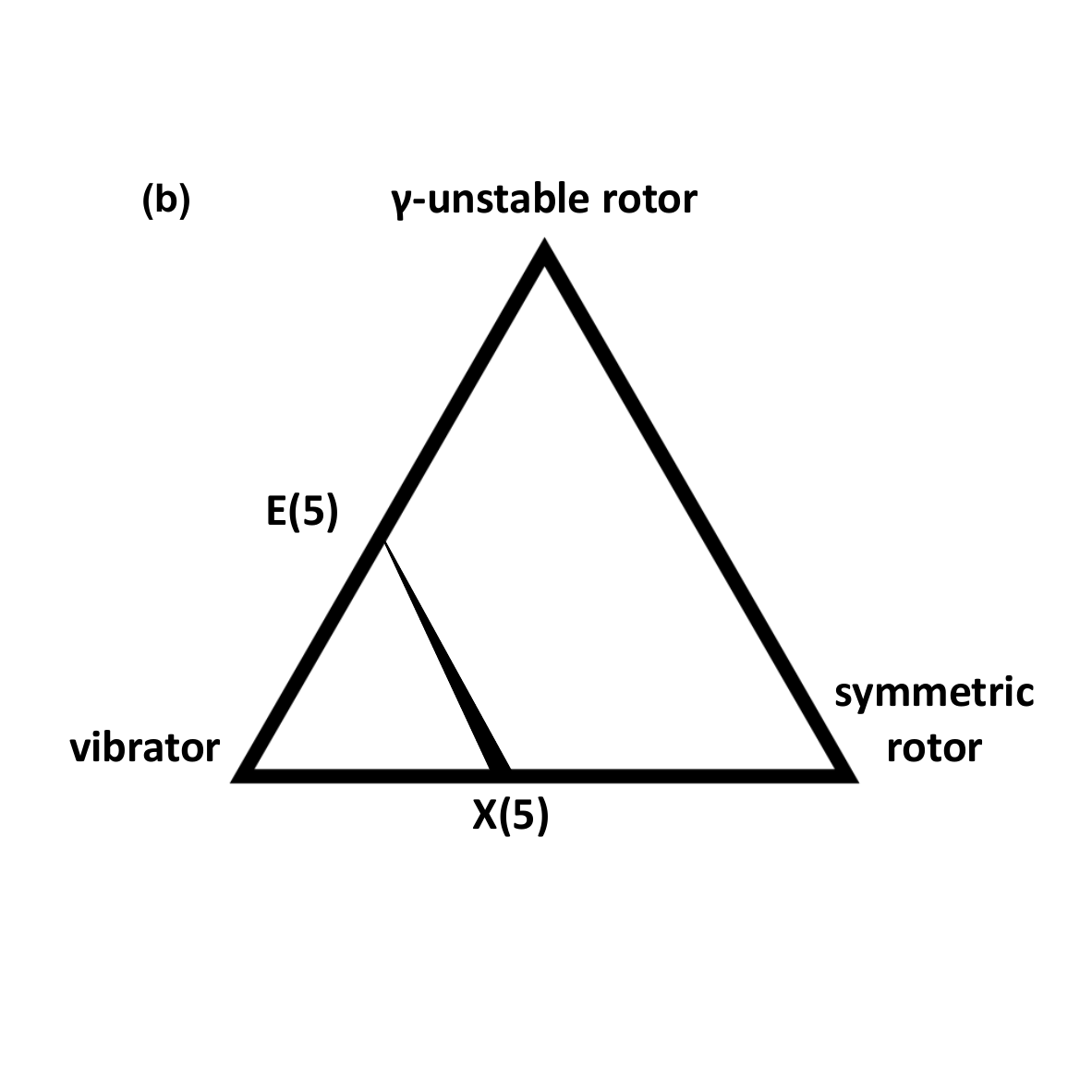}}

\caption{(a) Symmetry triangle of the IBM, depicting the narrow region surrounding the first order SPT separating the spherical and deformed phases, as well as the critical point of the second order SPT between U(5) and O(6) (see Fig. 2 of \cite{Feng1981} and Fig. 1 of \cite{Iachello1998}). See Sec. \ref{algebraic} for further discussion.
(b) Symmetry triangle of the collective model (see Fig. 3 of \cite{Zhang1997}), depicting the E(5) and X(5) critical point symmetries. See Sec. \ref{Bohr} for further discussion. }  
 \label{triangl1}

\end{figure*}

Phase transitions in the parameter space of the IBM have been first discussed in the framework of catastrophe theory \cite{Gilmore1981} in 1981 \cite{Feng1981}. A narrow region of first order phase transitions  separating the spherical [U(5)] and deformed [SU(3)] regions has been found, terminating at a second-order phase transition point lying between the spherical [U(5)] and $\gamma$-unstable [O(6)] regions (see Fig. \ref{triangl1}(a)). The possibility of a transition from prolate [SU(3)] to oblate 
[$\overline {\rm SU(3)}$] shapes through a $\gamma$-unstable point has been realized in 1996 \cite{LopezMoreno1996} (see Fig. 7 of \cite{LopezMoreno1996}). 

\begin{figure}[htb] 

\includegraphics[width=75mm]{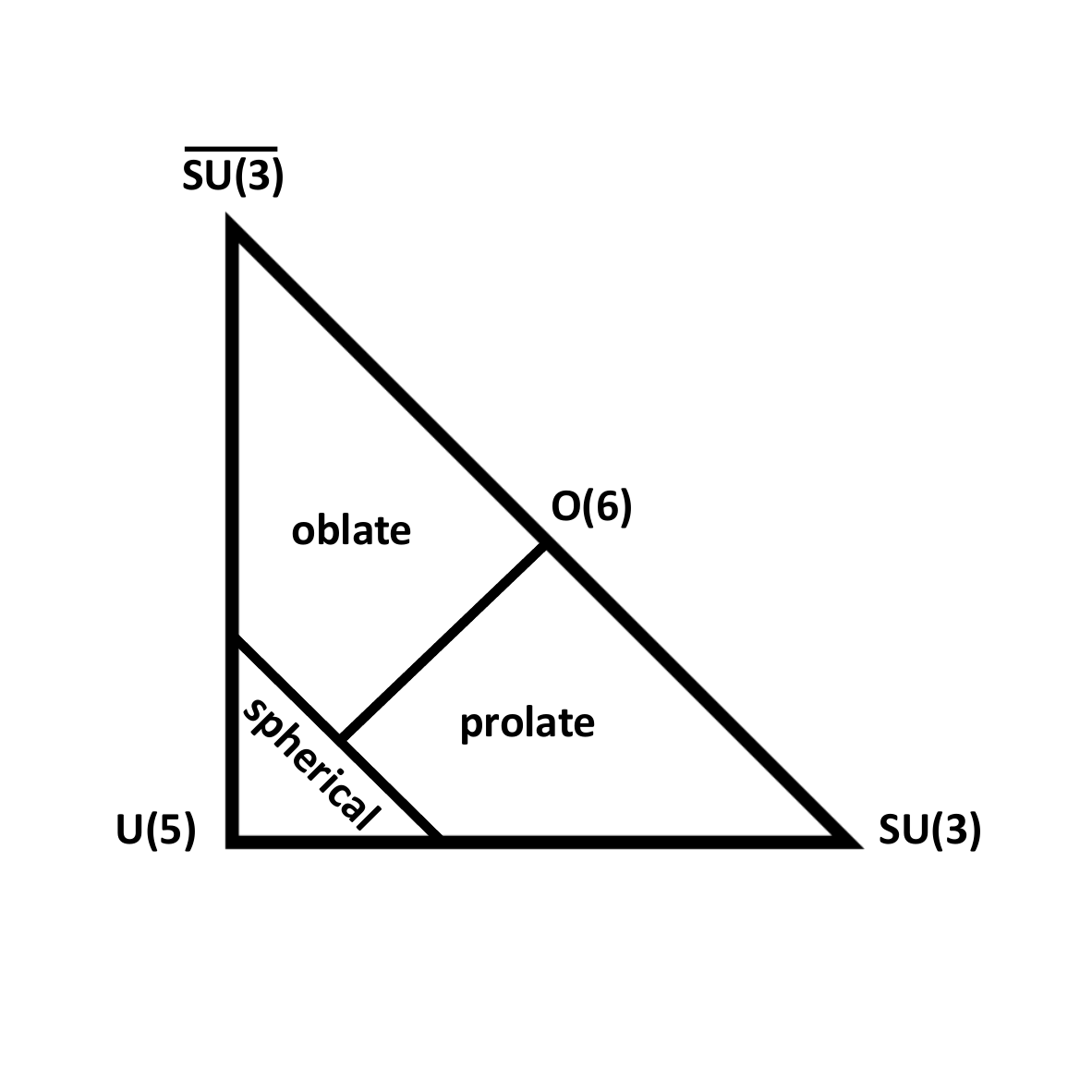}

\caption{Extended  symmetry triangle of the IBM, depicting the triple point separating the spherical, prolate, and oblate phases, as well as O(6) as the critical point of the phase transition from prolate to oblate shapes (see Fig. 4 of \cite{Jolie2001}, Fig. 1 of \cite{Jolie2002}, Fig. 1 of \cite{Warner2002}). See Sec. \ref{algebraic} for further discussion.
 }  
 \label{triangl2}

\end{figure}

A quantum phase transition from prolate to oblate shapes, having O(6) as its critical point, has been introduced in 2001 \cite{Jolie2001,Warner2002,Jolie2002} and tested against the data in 2003 \cite{Jolie2003}.  An order-parameter for this transition has been introduced in 2010 \cite{Bettermann2010}, in terms of quadrupole shape invariants \cite{Kumar1972,Cline1986}. This order parameter exhibits a peaking behavior at O(6) \cite{Bettermann2010}, thus characterizing the prolate to oblate transition as a first-order shape transition.   The prolate, oblate, and spherical regions meet at a single point, called the triple point, which represents a second-order shape/phase transition (SPT) \cite{Jolie2002,Warner2002} (see Fig. \ref{triangl2}). A detailed analysis of the IBM Hamiltonian in the consistent-$Q$ formalism \cite{Warner1983,Casten1988}, in which the same quadrupole operator is used in the Hamiltonian and for the transition operator in the $B(E2)$ transition rates, has been given in Ref. \cite{Pan2008}, performing numerical calculations taking advantage of the SU(3) Draayer-Akiyama basis \cite{Draayer1973,Akiyama1973}. 

In the above considerations, only one-body and two-body terms are taken into account in the IBM Hamiltonian, as it is done in the standrard IBM-1 model \cite{Iachello1987}, while advantage is taken of the parameter symmetry \cite{Shirokov1998} related to SU(3), according to which the quadrupole operator 
\begin{equation}\label{Qx}
Q = (d^\dagger s + s^\dagger d)^{(2)} +\chi (d^\dagger \tilde d)^{(2)} 
\end{equation}
satisfies the SU(3) commutation relations for both signs of the parameter $\chi=\mp \sqrt{7}/2$, with the negative (positive) sign corresponding to the prolate (oblate) shapes. The symbol SU(3) is used for the prolate case, while $\overline {\rm SU(3)}$ corresponds to the oblate shapes \cite{Shirokov1998}. The existence of this symmetry guarantees isospectrability \cite{Thiamova2006}, which implies that the prolate and oblate spectra are identical, thus the only way to distinguish between the two is the value of the  quadrupole moment of the $2_1^+$ state,  which should be negative (positive) for prolate (oblate) shapes \cite{Jolie2003}.   

Schematically, the parameter space of IBM-1 is depicted as a triangle, called Casten's triangle \cite{Casten1990}, with the three dynamical symmetries of IBM-1, U(5), SU(3), and O(6), occupying the three vertices of the triangle (see Fig. \ref{triangl1}(a)). In order to include $\overline {\rm SU(3)}$, a mirror image of the triangle along its U(5)-O(6) side is added, with 
$\overline {\rm SU(3)}$ being the mirror image of SU(3). Thus, schematically, O(6) appears midway between SU(3) and $\overline {\rm SU(3)}$ \cite{Jolie2001,Warner2002,Jolie2002} (see Fig. \ref{triangl2}). The robustness of the prolate to oblate SPT and the O(6) nature of the relevant critical point has been tested using an IBM Hamiltonian  with a linear dependence on the control parameter, with positive results \cite{Zhang2013}.   

An alternative path has been taken in the O(6) framework of IBM, in which three-body interactions (called the $QQQ$ interactions) have been found necessary for the construction of the rigid-rotor states of the Bohr Mottelson model \cite{Rowe2005,Thiamova2005,Thiamova2006}. The relevant Hamiltonian includes the second order Casimir operator of O(6) \cite{Iachello1987}, in addition to the two-body $QQ$ and three-body $QQQ$ interactions \cite{Thiamova2006}. The prolate to oblate transition is again found to be of the first order \cite{Thiamova2006}, while the triple point at which the prolate, oblate, and spherical shapes meet represents a second-order SPT with O(6) symmetry. The prolate and oblate shapes correspond to opposite signs of the cubic term, thus isospectrability occurs also in this case \cite{Thiamova2006}.   These findings have been corroborated by Ref. \cite{Fortunato2011}, in which the geometrical properties of an IBM Hamiltonian with cubic terms $QQQ$, in addition to the quadupole $QQ$ term and the vibrational $\hat n_d$ term, have been considered, with the relevant phase diagram constructed (see Fig. 3 of \cite{Fortunato2011}). A very tiny region of triaxiality is seen between the prolate and oblate phases (see Fig. 14 of \cite{Fortunato2011}).   
 
It should be mentioned that a parameter symmetry appears in the IBM-1 framework also in the O(6) case \cite{Shirokov1998}, related to the fact that the pairing operator used in the Hamiltonian bearing the DS O(6) can take the form  
\begin{equation} 
P= {1\over 2} (\tilde d \tilde d) + \xi (ss),
\end{equation}
with $\xi=\pm 1$. The symbols $P$ and O(6) are used for $\xi=-1$, while $\bar P$ and $\overline {\rm O(6)}$ are used for $\xi=+1$ \cite{VanIsacker1985,Shirokov1998}. 

It should be remembered that no triaxial shapes occur in IBM-1. The need for the inclusion of higher (third-order and fourth-order) terms in IBM-1, in order to include triaxial shapes, has been realized since 1981 \cite{VanIsacker1981,Heyde1984}. An alternative way for including triaxial shapes in the IBM framework is the use of IBM-2 \cite{Iachello1987}, in which distinction between bosons coming from correlated proton pairs and neutron pairs is made. Allowing protons to be described by SU(3) and neutrons by $\overline {\rm SU(3)}$ (or vice versa), the SU(3)$^*$ symmetry is obtained \cite{Dieperink1982,Dieperink1985,Walet1987}. SU(3) is used for valence nucleon particles  in the lower half of a nuclear shell, while $\overline {\rm SU(3)}$ is used for valence nucleon holes lying in the upper half of the nuclear shell, taking into account that in the IBM scheme the valence nucleons are counted from the nearest closed shell \cite{Iachello1987}. The use of oblate irreducible representations (irreps) in the upper half of the shell, used in the SU(3)$^*$ scheme, appears as a consequence of the short range nature of the nucleon-nucleon interaction, which leads to the preference for the highest weight irreps, in the framework of the proxy-SU(3) symmetry \cite{Bonatsos2017b,Martinou2021b}.
 
An important step forward has been  taken in 2012 in Ref. \cite{Zhang2012}, in which the most general IBM Hamiltonian including three-body terms has been solved analytically, rescaled in order to be expressed in terms of the second-order and third-order Casimir operators of SU(3), $\hat C_2$ and $\hat C_3$ respectively. The important difference occurring in this case is that the prolate to oblate transition becomes asymmetric, with the prolate spectra being different from the oblate spectra (see Fig. 2 of \cite{Zhang2012}), while the dynamical structure of the critical point is found to be similar but not identical to O(6), with a very tiny region of triaxiality occurring around the critical point for large boson numbers. The robustness of these findings has been tested \cite{Zhang2013} by using a transitional Hamiltonian with linear dependence on the control parameters, with positive results.  
 
A further important step has been taken in 2023 in Ref. \cite{Wang2023}, in which a vibrational term, $\hat n_d$, is added to the Hamiltonian containing the  second-order and third-order Casimir operators of SU(3). The prolate and oblate shapes are asymmetric, with the prolate shape exhibiting deformation on average twice that of the oblate shape. The ratio 
\begin{equation}
R_{3/2} = {E(2_3^+) \over E(2_2^+)} 
\end{equation}
has been suggested as an order parameter, since it acquires values close to unity on the prolate side, on which the $\beta$ and $\gamma$ bands are lying close to each other, while it becomes much larger on the oblate side (see Fig. 15 of \cite{Wang2023}), since on this side the bandhead of the $\gamma$-band, $2_2^+$, falls below the bandhead of the $\beta$-band, $0_2^+$ \cite{Wang2023}.  
 
\section{The Bohr collective model} \label{Bohr}

The collective model of Bohr and Mottelson \cite{Bohr1952,Bohr1953,Bohr1998a,Bohr1998b} has been very successful for many years in describing the properties of medium mass and heavy nuclei in terms of the collective variables $\beta$ and $\gamma$, corresponding to the departure from sphericity and to the departure from axiality respectively. 

Critical point symmetries (CPSs) in the framework of the Bohr Hamiltonian \cite{Bohr1952,Bohr1953,Bohr1998a,Bohr1998b} have been introduced by Iachello in 2000 \cite{Iachello2000}. The E(5) CPS \cite{Iachello2000} describes the second-order SPT from spherical to $\gamma$-unstable nuclei,  the X(5) CPS \cite{Iachello2001} corresponds to the first-order SPT from spherical to axially deformed nuclei, while the Y(5) CPS \cite{Iachello2003} describes the SPT from axial to triaxial nuclei. Additional CPSs have been proposed later, including the Z(5) CPS \cite{Bonatsos2004}, which corresponds to the SPT from prolate to oblate nuclei. 

In the E(5) and X(5) CPSs, an infinite square well potential is used in the $\beta$ variable, based on the expectation that for a SPT the potential should be flat, in order to allow for change of the shape at no energy expense. The potential in E(5) is independent from the $\gamma$ variable, thus allowing exact separation of variables in the relevant Schr\"{o}dinger equation \cite{Iachello2000}. In X(5) a sharp harmonic oscillator potential centered at $\gamma=0$ is used, in order to guarantee shapes close to prolate axial deformation. Separation of variables in this case is achieved only in an approximate way \cite{Iachello2001,Caprio2005}. 

The relation between symmetries and exact solvability of differential equations is well known \cite{Dresner1999,Hydon2000}. 
As a result, the E(5) CPS, which corresponds to an exact solution of Schr\"{o}dinger's differential equation, is indeed corresponding to the euclidean symmetry in five dimensions, Eu(5)\cite{Iachello2000,Caprio2007}, while no specific symmetry corresponding to X(5), which is related to an approximate solution of  Schr\"{o}dinger's differential equation, has been found up to date. 

It is instructive to place the E(5) and X(5) CPSs on the structural triangle for the geometric collective model \cite{Zhang1997}, the three vertices of which correspond to the vibrator, the axially symmetric rotor, and the $\gamma$-unstable rotor. Thus this triangle looks very similar to Casten's triangle, representing the parameter space of IBM-1, and having the U(5), SU(3) and O(6) DSs at the three corresponding vertices (see Fig. \ref{triangl1}(b)). The E(5) CPS is then represented by a point on the side of the collective structural triangle connecting the vibrator and the $\gamma$-unstable rotor, thus corresponding to the second-order SPT found in IBM-1 between the U(5) and O(6) DSs, while the X(5) CPS is represented by a point on the side of the collective structural triangle connecting the vibrator and the axially symmetric rotor, thus corresponding to the first-order SPT found in IBM-1 between the U(5) and SU(3) DSs. The fact that E(5) lies on the line connecting the U(5) and O(6) dynamical symmetries, which is known to be characterized by the O(5) symmetry \cite{Leviatan1986}, which is  a common subalgebra of U(5) and O(6), is in agreement with the fact that E(5) also possesses an O(5) subalgebra. 

It should be noticed that the spherical and axially symmetric rotor phases are not separated by a single line, but by a narrow region, which becomes more and more narrow as the vibrator to $\gamma$-unstable rotor side is approached, becoming, when reaching it, the single point representing the E(5) CPS. This fact is known in the IBM framework since the seminal paper of 1981 \cite{Feng1981}, and has been pointed out as a region of shape coexistence in Ref. \cite{Iachello1998}, while its borders have been worked out in Refs. \cite{Zamfir2002,McCutchan2006}. Taking into account only $n_\gamma=0$ bands in X(5), one can see that the remaining bands can be accommodated within the Eu(5) symmetry, thus building a Eu(5) bridge between the E(5) and X(5) CPSs \cite{Zhang2014a}. The triple point at which the spherical, prolate, and oblate shapes meet, is then also characterized by the Eu(5) symmetry \cite{Zhang2014b} (see Fig. 1 of \cite{Zhang2014b}). 

Exact separation of variables in the X(5) framework can be achieved by freezing the $\gamma$ variable at $\gamma=0$, in which case the X(3) CPS \cite{Bonatsos2006} is obtained. To the best of our knowledge, no symmetry associated to X(3) has been identified up to date, despite the fact that exact separability of variables indicates that its existence could be possible. 

In the Z(5) CPS \cite{Bonatsos2004}, an infinite square well potential is used in the $\beta$ variable, while in the $\gamma$ variable a steep harmonic oscillator centered around $\gamma=30^{\rm o}$ is used, midway between the prolate $\gamma=0$ and oblate $\gamma=60^{\rm o}$ shapes. Separation of variables is achieved in an approximate way 
\cite{Bonatsos2004}, parallel to the one used in the X(5) case \cite{Iachello2001,Caprio2005}.  Exact separation of variables in the Z(5) framework can be achieved by freezing the $\gamma$ variable at $\gamma=30^{\rm o}$, in which case the Z(4) CPS \cite{Bonatsos2005} is obtained. In this case, a partial identification with a specific symmetry exists, since the ground state band of Z(4) has been found to be identical to the ground state band of the euclidean symmetry in four dimensions, E(4) \cite{Bonatsos2005}. 

An important advantage of the above-mentioned CPSs is that they provide parameter-independent (up to overall scale factors) predictions for spectra and $B(E2)$ transition rates, therefore stringent tests of their validity against the data are possible, as one can see in the review articles \cite{Casten2006,Casten2007,Casten2009,Cejnar2010}. 

Prolate-oblate symmetry is present in the Z(5) and Z(4) CPSs. However, prolate-oblate asymmetry in the frameowork of the Bohr Hamlitonian has been suggested already in 1974 \cite{Rohozinski1974}, by adding a $\beta^3 \cos 3\gamma$ to the $\beta$-potential, which is a function of $\beta^2$, the Davidson potential for example \cite{Rohozinski1974}.
     
Several variations of the Z(5) solution exist in the literature, taking advantage of its approximate solvability. In these variations the infinite well potential in the 
$\beta$ variable is replaced by a sextic potential \cite{Budaca2016}, a Kratzer potential \cite{Hammad2021b}, a Morse potential \cite{Hammad2023}, a Tietz-Hua potential 
\cite{Hammad2023}, or a multi-parameter exponential type potential \cite{Hammad2023}. In Refs. \cite{Hammad2021b,Hammad2023} a conformable fractional Bohr Hamiltonian is used, which is a generalization of the Bohr Hamiltonian in which the usual derivatives are replaced by conformable fractional derivatives \cite{Khalil2014}, which allow for fractional orders of derivatives, while preserving  the familiar properties of usual derivatives. Conformable fractional derivatives, a special case of fractional derivatives \cite{Miller1993,Podlubny1999,Herrmann2011} introduced in the study of critical point symmetries by Hammad \cite{Hammad2021a}, contain an extra parameter, the order of the derivative, thus being able to approach closer to the critical point than usual derivatives. 

Several variations of the Z(4) solution exist in the literature, taking advantage of its exact solvability. In these variations the infinite well potential in the $\beta$ variable is replaced by a sextic potential \cite{Buganu2015,Budaca2016}, a Davidson potential \cite{Yigitoglu2018}, a Kratzer potential \cite{Heydari2018},  a Davidson potential with a deformation-dependent mass \cite{Buganu2017}, or a Kratzer potential with a deformation-dependent mass \cite{AitElKorchi2021,AitElKorchi2022}. The deformation-dependent mass formalism \cite{Bonatsos2011,Bonatsos2013}, based on supersymmetric quantum mechanics \cite{Cooper1995,Cooper2001}, reduces the rate of increase of the nuclear moment of inertia with increasing deformation, thus removing a major drawback of the Bohr Hamiltonian \cite{Bohr1952}. 
 
Reviews of special solutions of the Bohr Hamiltonian for various potentials related to SPTs have been given in Refs. \cite{Fortunato2005,Buganu2016}.  

The contents of the present section and the previous one reveal a serious contradiction between the description of the CPS of the prolate to oblate transition in the IBM and Bohr frameworks. In the IBM framework the critical point is characterized by the O(6) symmetry, which is $\gamma$-unstable, while in the Bohr framework the critical point is characterized by the Z(5) solution, which is nearly $\gamma$-rigid, possessing a maximally triaxial shape ($\gamma=30^{\rm o}$). This apparent contradiction has been resolved since 1987 by Otsuka and Sugita \cite{Otsuka1987}, who have proved the equivalence between $\gamma$-instability and triaxiality in the IBM framework for finite boson systems, which is the case for realistic nuclei. The equivalence between $\gamma$-instability and rigid triaxiality for finite boson numbers has been corroborated for the ground states by Cohen \cite{Cohen1988}. Since we are interested in ground state SPTs, this equivalence suffices.   
 
\section{Mean field models} \label{meanfield}
                                                                                                                                                                                                                                                                                                                                                                                                                                                                                                                                                                                                                                                                                                                                                                                                             Self-consistent mean field models have been used in nuclear structure for a long time \cite{Bender2003}, starting with non-relativistic Skyrme \cite{Skyrme1956,Skyrme1959,Erler2011} and Gogny \cite{Gogny1973,Gogny1975,Delaroche2010} energy density functionals and evolving towards relativistic energy density functionals \cite{Vretenar2005,Niksic2011}. 

The differences between prolate and oblate shapes have been pointed out  by Kumar already in 1970 \cite{Kumar1970}, based on calculations within the pairing plus quadrupole (PPQ) model \cite{Baranger1968,Kumar1968} in the W-Os-Pt region. In particular, it has been suggested that the energy difference between the bandhead of the $\gamma$-band and the $L=4$ member of the ground state band, $E(2_2^+) -E(4_1^+)$, should be positive (negative) for prolate (oblate) nuclei (see Fig. 1 of \cite{Kumar1970}). Furthermore, a transition between spherical-prolate-oblate-spherical shapes has been suggested in the rare earth region with $Z=50-82$ and $N=82-126$ \cite{Kumar1972} (see Fig. 1 of \cite{Kumar1972}).  The $E(2_2^+) -E(4_1^+)$ systematics have been extended to the fp shell, through a Hartree-Fock-Bogoliubov (HFB) calculation using the generator coordinate method \cite{Castel1976} (see Table I of \cite{Castel1976}).

A next step has been taken by using the Woods-Saxon potential and the modified harmonic oscillator (Nilsson) potential in Strutinsky plus BCS calculations for determining potential energy surfaces (PESs) in the Pt-Hg-Pb region \cite{Bengtsson1987,Nazarewicz1993} (see Fig. 1 of \cite{Bengtsson1987} and Fig. 3 of \cite{Nazarewicz1993}). Shape coexistence \cite{Heyde1983,Wood1992,Heyde2011,Bonatsos2023c} of low-lying prolate and oblate $0^+$ states, one of them being the ground state and the other one lying close in energy above it, has been identified \cite{Bengtsson1987,Nazarewicz1993}. In addition, a prolate to oblate transition has been identified in the Os series of isotopes \cite{Nazarewicz1990} (see Table 1 of \cite{Nazarewicz1990}). Furthermore, oblate ground states in light nuclei have been predicted using Nilsson diagrams based on realistic Woods-Saxon potentials \cite{Hamamoto2014}.    

The advent of RMF theory allowed for calculations exhibiting prolate and oblate shapes in the Pt-Hg-Pb region \cite{Sharma1992,Yoshida1994,Patra1994}. The sensitive dependence of the results on the set of input parameters has been pointed out \cite{Heyde1996,Takigawa1996}. Potential energy curves (PECs) have been calculated for various series of isotopes \cite{Fossion2006,Wang2012}, allowing for the observation of the sudden transition of the ground state from a prolate to an oblate shape along some series of isotopes \cite{Fossion2006,Wang2012} (see Figs. 1-8,10,13 of \cite{Fossion2006}). The need to include the $\gamma$ degree of freedom in the RMF calculations had already been realized \cite{Fossion2006}.  

At about the same time, PECs exhibiting competing prolate and oblate minima in the Yb-Pt region have also been calculated in the non-relativistic mean field framework, using the self-consistent axially-deformed Hartree-Fock (HF) \cite{Stevenson2005} and the self-consistent axially-symmetric Skyrme HF plus BCS \cite{Sarriguren2008} approaches (see Fig. 2 of \cite{Stevenson2005} and Fig. 2 of \cite{Sarriguren2008}). Soon thereafter the non-relativistic mean field approaches were extended to the calculation of PESs, using a self-consistent HFB approach with the Gogny D1S and the Skyrme SLy4 interactions \cite{Robledo2009}, as well as with the Gogny D1N and D1M parametrizations \cite{RodriguezGuzman2010} (see Figs. 3 and 5 of \cite{Robledo2009}, as well as Figs. 2-4 of \cite{RodriguezGuzman2010}). 

A major step forward has been taken in 2008 by Nomura \textit{et al.} \cite{Nomura2008,Nomura2009,Nomura2010}, through the determination of the parameters of the IBM Hamiltonian by adjusting its PES to agree with the PES predicted by RMF calculations, thus making affordable the calculation of spectra and B(E2) transition rates of series of isotopes. Level schemes have been calculated in the W-Pt region using the Gogny D1S interaction \cite{Nomura2011a,Nomura2011b}), as well as in the Yb-Pt region using the Gogny D1M interaction \cite{Nomura2011d} (for level schemes see Fig. 5 of \cite{Nomura2011a} and Fig. 5 of \cite{Nomura2011b}). The method has been recently extended to the study of prolate to oblate shape phase transitions in odd-mass nuclei \cite{Nomura2018} in the 
Os-Ir-Pt region using the D1M interaction and the Interacting Boson Fermion Model(IBFM) \cite{Iachello1991}.

PESs and level schemes for series of isotopes in the Er-Pt region have also been calculated recently \cite{Yang2021a} by using a five-dimensional quadrupole collective Hamiltonian with parameters determined from covariant density functional theory (CDFT) calculations using the PC-PK1 energy density functional (see Figs. 1-6 of \cite{Yang2021a} for PESs and Fig. 9 for level schemes).  

\section{The shell model} \label{shellmodel}

The nuclear shell model \cite{Heyde1990,Talmi1993} has been the fundamental microscopic theory of nuclear structure since its introduction \cite{Mayer1948,Mayer1949,Haxel1949,Mayer1955}. 
 
Shell model calculations in the $A\approx 110$ region for the $N=66$ series of isotones \cite{Xu2002} have indicated a transition from prolate (Se-Zr) through triaxial (Mo, Ru) to oblate (Pd) shapes. 

Furthermore, shell model calculations have been performed for the $N=Z$ nuclei in the $fpg$ shell \cite{Kaneko2004}, in relation to the search for shape coexistence of prolate and oblate shapes, one of them characterizing the ground state and the other a low-lying $0^+$ state.  The importance of four-particle four-hole (4p-4h) excitations for the creation of oblate states in $^{68}$Se and $^{72}$Kr has been pointed out, while this is not the case in $^{60}$Zn and $^{64}$Ge.    
 
It should be noticed that a different kind of prolate to oblate  transition has been proposed within the projected shell model in the $^{190}$W$_{116}$ nucleus, namely the sudden change of the Yrast band from prolate to oblate shapes at angular momentum $L=10$ \cite{Sun2008}. We are not going to consider further this type of structural change in the present article. 
 
Recent shell model studies in the Sn-Sm region within the $N=50-82$ shell have demonstrated the need to include the $2f_{7/2}$ orbital in projected HFB plus generator coordinate method calculations in the model space consisting of the $1g_{9/2}$, $1g_{7/2}$, $2d_{5/2}$, $2d_{3/2}$, $3s_{1/2}$, $1h_{11/2}$ orbitals \cite{Kaneko2021,Kaneko2023}. It has been found that the neutron partner orbitals ($1h_{11/2}$, $2f_{7/2}$) are crucial for describing the prolate to oblate SPT occurring at $N=76$. The need to take into account the $\Delta j=2$ orbitals ($1h_{11/2}$, $2f_{7/2}$) within the $sdg$ shell, and not only the intruder orbital $1h_{11/2}$, in order to correctly reproduce the nuclear properties \cite{Kaneko2021,Kaneko2023}, is a hallmark of the approximate quasi-SU(3) symmetry, initially proposed in 1995 by Zuker \textit{et al.} \cite{Zuker1995,Zuker2015} in the case of the $pf$ shell, in which the addition of the $\Delta j=2$ orbitals ($2d_{5/2}$, $3s_{1/2}$) has been found necessary for  building up the appropriate (high enough) collectivity \cite{Zuker1995,Zuker2015}.    
 
Another approximation scheme allowing the use of the SU(3) symmetry in medium mass and heavy nuclei beyond the $sd$ shell, in which the microscopic Elliott SU(3)symmetry \cite{Elliott1958a,Elliott1958b,Elliott1963,Elliott1968} is destroyed by the spin-orbit interaction, is the proxy-SU(3) symmetry, introduced by Bonatsos \textit{et al.} in 2017 \cite{Bonatsos2017a,Bonatsos2017b,Bonatsos2017c,Bonatsos2023a}. The connection of the proxy-SU(3) symmetry to the shell model has been clarified in \cite{Martinou2020} (see Table 7 of \cite{Martinou2020}), while its connection to the Nilsson model has been demonstrated in \cite{Bonatsos2017a,Bonatsos2020b}. The microscopic foundation of proxy-SU(3) symmetry is based on pairs of Nilsson orbitals having identical angular momenta and spin, while they differ by one oscillator quantum in the $z$-direction. These pairs have been found empirically to correspond to maximum proton-neutron interaction \cite{Cakirli2010,Cakirli2013,Casten2016}, while in parallel they exhibit maximal spatial overlap \cite{Bonatsos2013b}. Using the well-established correspondence between SU(3) and the rigid rotor \cite{Castanos1988}, the proxy-SU(3) symmetry provides parameter-free predictions for the collective $\beta$ and $\gamma$ variables, based on the highest weight SU(3) irreducible representations (irreps) for the valence protons and the valence neutrons of a given nucleus \cite{Bonatsos2017b,Bonatsos2017c,Martinou2021b}. The highest weight irreps are identical to the irreps possessing the highest eigenvalue of the second order Casimir operator of SU(3) up to the middle of each shell, but they are different beyond the middle of the shell (see Table I of \cite{Bonatsos2017b} and/or Table I of \cite{Sarantopoulou2017}), the difference being rooted in the short-range nature of the nucleon-nucleon interaction, which favors the most symmetric irreps allowed by the Pauli principle \cite{Martinou2021b} (see Tables 1-8 of \cite{Martinou2021b}). In the rare earth region between $Z=50-82$ and $N=82-126$, the proxy SU(3) symmetry predicts a prolate to oblate transition around $N=114$ (see Table II of \cite{Bonatsos2017b}), while in the region $Z=50-82$ and $N=50-82$ a prolate to oblate transition around $N=72$ is predicted (see Table III of \cite{Bonatsos2017b}). A by-product of proxy-SU(3) symmetry and the dominance of the highest weight irreps is the resolution \cite{Bonatsos2017b,Bonatsos2017c,Martinou2021b,Sarantopoulou2017} of the long standing problem of the dominance of prolate over oblate shapes \cite{Hamamoto2009,Hamamoto2012} in the ground states of even-even nuclei. Another by-product of the proxy-SU(3) symmetry is the dual shell mechanism \cite{Martinou2021a}, which predicts that shape coexistence can occur only on specific islands of the nuclear chart \cite{Martinou2023,Bonatsos2023b,Bonatsos2023c}, recently corroborated through CDFT calculations \cite{Bonatsos2022a,Bonatsos2022b}.
 
Yet another approximation scheme allowing the use of the SU(3) symmetry in medium-mass and heavy nuclei beyond the $sd$ shell is the pseudo-SU(3) scheme \cite{Arima1969,Hecht1969,RatnaRaju1973,Draayer1982,Draayer1983,Draayer1984,Castanos1987,Bahri1992,Ginocchio1997}, in which the SU(3) symmetry is approximately restored for the normal parity orbitals remaining in a nuclear shell after the sinking of the orbital possessing the highest angular momentum into the shell  below, because of the spin-orbit interaction. The intruder orbital, invading from the shell above, possessing the opposite parity, and therefore also called the abnormal parity orbital, is treated separately by shell model methods, without participating in the formation of the pseudo-SU(3) symmetry. Despite the fact that pseudo-SU(3) and proxy-SU(3) make different approximations in order to restore the SU(3) symmetry in the presence of the spin-orbit interaction, and are based on different unitary transformations within the shell model space (see \cite{Castanos1992a,Castanos1992b,Castanos1994} for the unitary transformation used in pseudo-SU(3) and \cite{Martinou2020} for the one used in proxy-SU(3)), it turns out that they provide very similar predictions for the $\beta$ and $\gamma$ collective variables, as well as for the points of transition from prolate to oblate shapes, as shown in Ref. \cite{Bonatsos2020a}, provided that the highest weight irreps are used in both cases.  
 
The various applications of SU(3) symmetry in atomic nuclei have been recently reviewed by  Kota \cite{Kota2020}, who has recently extended the proxy-SU(3) symmetry to a proxy-SU(4) symmetry \cite{Kota2023} in the $A=60$-90 region, in which the valence protons and valence neutrons occupy the same major shell, so that the spin-isospin Wigner SU(4) symmetry \cite{Wigner1937} becomes important \cite{Kota2017}.

\section{Statistical methods} \label{statistical} 

Some recent applications of statistical methods in nuclear structure are briefly reviewed in this section. 

\subsection{Entanglement entropy} 

Quantum entanglement is an effect which occurs in systems of particles that interact in such a way that the quantum state of each particle cannot be described independently of the state of the others, even in the case in which they are separated by a large distance. Its discussion started with the \textsl{Gedankenexperiment} (thought experiment) of Einstein, Podolsky, and Rosen in 1935 \cite{Einstein1935}.

A measure suitable for entanglement is the von Neumann entropy \cite{Vedral1997,Vedral2014,Cao2006}, which has been recently used \cite{Jafarizadeh2022} as an order parameter in the study of quantum SPTs in nuclei in the framework of the IBM and IBFM models. In particular, the study of nuclei in the transitional region from U(5) to O(6) showed that no entanglement between the $s$ and $d$ bosons exists in the U(5) limit, while maximum entanglement between the $s$ and $d$ bosons is seen in the O(6) limit \cite{Jafarizadeh2022}. 
 
\subsection{Fluctuations of shape variables}
 
The shape variables $\beta$ and $\gamma$ are known \cite{Kumar1972b,Cline1986} to be determinable in a model-independent way from the quadratic and cubic invariants 
$q_2=(QQ)^{(0)}$ and $q_3=(QQQ)^{(0)}$ of the quadrupole operator $Q$. In particular, $\beta^2$ is proportional to $q_2$, while $\cos 3\gamma$  is proportional to 
$q_3/ q_2^{3/2}$ \cite{Elliott1986}, while higher order invariants have also been used \cite{Werner2001,Werner2005}. 

Through the use of higher order invariants, recent studies \cite{Poves2020} of the fluctuations of the $\beta$ and $\gamma$ variables within the configuration-interaction shell model for $pf$ shell nuclei indicate that $\beta$ is often characterized by a non-negligible degree of softness, while $\gamma$ usually has large fluctuations, making its value not meaningful \cite{Poves2020}. Particularly large fluctuations appear especially in the case of doubly magic nuclei, rendering questionable the characterization of their shape as "spherical" \cite{Poves2020}. 
 
\section{Predictions for prolate to oblate transition in specific nuclei} \label{po}

In this section, the specific predictions made by the theoretical methods reviewed in sections \ref{algebraic}-\ref{statistical} are summarized, arranged in series of isotopes. 

\subsection{The Pt (${\bf Z=78}$) isotopes} \label{Pt}

The proxy-SU(3) symmetry predicts a prolate to oblate transition between $N=114$ (prolate) and $N=116$ (oblate) (see Table I of \cite{Bonatsos2017b}). It should be noticed that the proxy-SU(3) irreps in the proxy $pfh$ (82-126) shell start being oblate at 34 valence neutrons, i.e., at $N=116$ (see Table I of \cite{Bonatsos2017b}). 

The pseudo-SU(3) symmetry, when the highest weight irreps are used, predicts a prolate to oblate transition between $N=112$ (prolate) and $N=114$ (oblate) (see Table 3 of \cite{Bonatsos2020a}). The agreement with proxy-SU(3) is very good, remembering that the intruder orbitals are ignored in the present application of the pseudo-SU(3) approach. 

The parameter-free predictions (up to an overall scale factor) of the Z(5) CPS corresponding to the prolate to oblate SPT \cite{Bonatsos2004}, which is an approximate solution of the Bohr Hamiltonian, give best agreement to the data for $N=116$ ($^{194}$Pt), as seen in Table 3 of \cite{Bonatsos2004}.  

It should be noticed that all three of the above models provide parameter-free predictions for the prolate to oblate transition point. The fact that their predictions are in very good agreement, despite the different origins of each model and the different approximations made in each of them, adds credibility to their predictions.

Various results obtained in the framework of the IBM corroborate the above predictions. The IBM calculations of Ref. \cite{Jolie2003}, using the standard IBM Hamiltonian including one- and two-body terms,  predict oblate shapes at $N=116$, 118 (see Table I of \cite{Jolie2003}). The IBM calculations of Ref. \cite{Zhang2012},  in which higher-order interactions are included, predict oblate shapes at $N=114$-120 (see Fig. 9 of \cite{Zhang2012}). The IBM calculations of Ref. \cite{Wang2023}, in which again higher-order interactions are included, are also compatible with oblate shapes at $N=114$-120 (see Fig. 15 of \cite{Wang2023} for the experimental data of the $R_{3/2}= E(2_3^+)/E(2_2^+)$ ratio corroborating this result). 

Several results obtained by non-relativistic mean-field methods also corroborate the above predictions. HF calculations \cite{Stevenson2005} suggest $N=116$ as the critical point (see Fig. 2 of \cite{Stevenson2005}). HFB calculations with the Skyrme Gogny D1S and Sly4 interactions \cite{Robledo2009} also suggest $N=116$ as the critical point, with the critical region exhibiting a $\gamma$-soft behavior (see Figs. 2, 3, 5 and Table 1 of \cite{Robledo2009}).  

In the RMF realm, early calculations using the NL3 interaction \cite{Fossion2006} suggested oblate shapes at $N=116$, 118 (see Fig. 9 of \cite{Fossion2006}), with the reservation that the $\gamma$ degree of freedom has also to be taken into account. Calculations obtaining the complete PESs using the Gogny D1S interaction \cite{Nomura2011a} and the D1M interaction \cite{Nomura2011d}, and predicting spectra and B(E2) transition rates through an IBM Hamiltonian with its parameters fitted to the RMF PES, conclude that $N=116$ can be considered as the prolate to oblate transition point (see Fig. 1 of \cite{Nomura2011a} and Fig. 1 of \cite{Nomura2011d} for the relevant PES, as well as Table I of \cite{Nomura2011d} for the relevant IBM parameters). 

Results compatible with a prolate to oblate transition between $N=114$ and $N=116$ have also been obtained using a 5-dimensional quadrupole collective Hamiltonian with parameters determined from the PC-PK1 energy density functional \cite{Yang2021a} (see Fig. 6 of \cite{Yang2021a} for the relevant PES). 

On the empirical front, the energy splitting $E(2_\gamma^+)-E(4_1^+)$, proposed by Kumar \cite{Kumar1970}, suggests oblate shapes at $N=114$-118 (see Fig. 1 of \cite{Kumar1970}). Experimental results for $^{192}$Pt$_{114}$  \cite{Oktem2007} suggest that a prolate to oblate transition takes place in this region. 

In summary, theoretical predictions and data agree on a prolate to oblate transition at $N=114$-116. 

 \subsection{The Os (${\bf Z=76}$) isotopes}
 
The proxy-SU(3) symmetry again predicts, as in Pt, a prolate to oblate transition between $N=114$ (prolate) and $N=116$ (oblate) (see Table I of \cite{Bonatsos2017b}). 

Also, the pseudo-SU(3) symmetry, when the highest weight irreps are used, again predicts, as in Pt, a prolate to oblate transition between $N=112$ (prolate) and $N=114$ (oblate) (see Table 3 of \cite{Bonatsos2020a}).

The IBM calculations of Ref. \cite{Jolie2003}, using the standard IBM Hamiltonian including one- and two-body terms,  predict prolate shapes at $N=112$-116 (see Table I of \cite{Jolie2003}). The IBM calculations of Ref. \cite{Zhang2012},  in which higher-order interactions are included, predict prolate shapes at $N=112$-114 (see Fig. 9 of \cite{Zhang2012}).

Early calculations in the shell correction approach using a Woods-Saxon potential and the monopole pairing \cite{Nazarewicz1990} predict prolate shapes up to $N=114$ and oblate shapes from $N=116$ onward (see Table 1 of \cite{Nazarewicz1990}).   

Non-relativistic mean-field approaches also corroborate the above predictions. HF calculations \cite{Stevenson2005} suggest $N=116$ as the critical point (see Fig. 2 of \cite{Stevenson2005}). Skyrme HF plus BCS calculations \cite{Sarriguren2008} also suggest that the prolate to oblate transition takes place between $N=116$ and 118 (see Fig. 2 of \cite{Sarriguren2008} for the relevant PECs). HFB calculations with the Skyrme Gogny D1S and Sly4 interactions \cite{Robledo2009} also suggest $N=116$ as the critical point, with the critical region exhibiting a $\gamma$-soft behavior (see Figs. 2, 3, 5 and Table 1 of \cite{Robledo2009}).  

Early RMF calculations using the NL3 interaction \cite{Fossion2006} suggested prolate shapes at $N=112$-116 (see Fig. 9 of \cite{Fossion2006}). Calculations obtaining the complete potential PESs using the Gogny D1S interaction \cite{Nomura2011b} and the D1M interaction \cite{Nomura2011d}  conclude that $N=116$ can be considered as the prolate to oblate transition point (see Fig. 1 of \cite{Nomura2011b} and Fig. 1 of \cite{Nomura2011d} for the relevant PES, as well as Table I of \cite{Nomura2011d} for the relevant IBM parameters).  

Results compatible with a prolate to oblate transition between $N=114$ and $N=116$ have also been obtained using a 5-dimensional quadrupole collective Hamiltonian with parameters determined from the PC-PK1 energy density functional \cite{Yang2021a} (see Fig. 5 of \cite{Yang2021a} for the relevant PES). 

On the empirical front, the energy splitting $E(2_\gamma^+)-E(4_1^+)$ proposed by Kumar \cite{Kumar1970} suggests prolate shapes at $110$-114 (see Fig. 1 of \cite{Kumar1970}) and an oblate shape at $N=118$. Early experimental results suggest a prolate to oblate transitional character for  $^{192}$Os$_{116}$  \cite{Casten1978}, and an oblate shape for $^{194}$Os$_{116}$\cite{Casten1978,Wheldon2000}, while more recent experiments find an oblate shape in  $^{198}$Os$_{122}$ \cite{Podolyak2009}, and suggest a prolate to oblate transition at  $^{190}$Os$_{116}$ \cite{John2014} and oblate shapes in $^{192-198}$Os$_{116-122}$ \cite{John2014} (see Fig. 3 of \cite{John2014}).

In summary, theoretical predictions and data are consistent with a prolate to oblate transition around $N=116$. 
 
\subsection{The Yb-Hf-W (${\bf{Z=70}}$-74) isotopes} 
 
The proxy-SU(3) symmetry predicts in Hf and W a prolate to oblate transition, with the oblate region starting at $N=118$ and $N=116$ respectively (see Table I of \cite{Bonatsos2017b}). 

The pseudo-SU(3) symmetry, when the highest weight irreps are used, predicts in Hf and W a prolate to oblate transition, with the oblate region starting at $N=116$ and $N=114$ respectively (see Table 3 of \cite{Bonatsos2020a}).
 
The IBM calculations of Ref. \cite{Jolie2003}, using the standard IBM Hamiltonian including one- and two-body terms,  predict prolate shapes at $N=108$-112 for W and at $N=108$ for Hf (see Table I of \cite{Jolie2003}). The same predictions are made by the IBM calculations of Ref. \cite{Zhang2012},  in which higher-order interactions are included (see Fig. 9 of \cite{Zhang2012}). 
 
Non-relativistic mean-field approaches provide uniform predictions for the Yb, Hf, and W series of isotopes. HF calculations \cite{Stevenson2005} suggest $N=116$ as the critical point (see Fig. 2 of \cite{Stevenson2005}). Skyrme HF plus BCS calculations \cite{Sarriguren2008} also suggest that the prolate to oblate transition takes place between $N=116$ and 118 (see Fig. 2 of \cite{Sarriguren2008} for the relevant PECs). HFB calculations with the Skyrme Gogny D1S and Sly4 interactions \cite{Robledo2009} also suggest $N=116$ as the critical point, with the transition being sharp in Yb and Hf, while in W the critical region exhibits a $\gamma$-soft behavior (see Figs. 2, 3, 5 and Table 1 of \cite{Robledo2009}).   
 
Early RMF calculations using the NL3 interaction \cite{Fossion2006} suggested prolate shapes at $N=108$-112 for W and at $N=108$ for Hf (see Fig. 9 of \cite{Fossion2006}). Calculations obtaining the complete PESs using the Gogny D1S interaction \cite{Nomura2011b} and the D1M interaction \cite{Nomura2011d}  conclude that $N=116$ can be considered as the prolate to oblate transition point in Yb, Hf, and W (see Fig. 1 of \cite{Nomura2011b} and Fig. 1 of \cite{Nomura2011d} for the relevant PES, as well as Table I of \cite{Nomura2011d} for the relevant IBM parameters).  

Results compatible with a prolate to oblate transition between $N=114$ and $N=116$ in Yb, Hf, and W, as well as in Er, have also been obtained using a 5-dimensional quadrupole collective Hamiltonian with parameters determined from the PC-PK1 energy density functional \cite{Yang2021a} (see Figs. 1-4 of \cite{Yang2021a} for the relevant PES). 
 
On the empirical front, experimental results for $^{190}$W$_{116}$ suggest it as the point of a transition from prolate to oblate shapes, having at the same time maximal $\gamma$-softness \cite{Alkhomashi2009}.  

In summary, theoretical predictions and data are consistent with a prolate to oblate transition around $N=116$. 

\subsection{The Hg (${\bf Z=80}$) isotopes} \label{Hg}

The proxy-SU(3) symmetry predicts in Hg a prolate to oblate transition, with the oblate region starting at $N=116$ (see Table I of \cite{Bonatsos2017b}).  

The IBM calculations of Ref. \cite{Jolie2003}, using the standard IBM Hamiltonian including one- and two-body terms,  predict oblate shapes at $N=118$, 120 (see Table I of \cite{Jolie2003}). The IBM calculations of Ref. \cite{Wang2023}, in which higher-order interactions are included, are also compatible with oblate shapes at $N=118$, 120 (see Fig. 15 of \cite{Wang2023} for the experimental data for the $R_{3/2}= E(2_3^+)/E(2_2^+)$ ratio corroborating this result). 

Early RMF calculations using the NL3 interaction \cite{Fossion2006} suggested oblate shapes at $N=118$, 120 (see Fig. 9 of \cite{Fossion2006}), with the reservation that the $\gamma$ degree of freedom has also to be taken into account.

On the empirical front, experimental results \cite{Bockisch1979} for $^{200-204}$Hg$_{120-124}$ suggest oblate shapes for these isotopes. 

In summary, theoretical predictions and data are consistent with a prolate to oblate transition around $N=116$. 

\subsection{The ${\bf Z=50}$-68 region} \label{Z5082}
 
Recent (2021) shell model calculations using the quasi-SU(3) symmetry in the $Z=50$-62 region suggest \cite{Kaneko2021} prolate shapes in the ground state for $N\leq 76$ and oblate shapes above it, emphasizing the crucial role played by the $2f_{7/2}$ orbital in obtaining this result. More recent (2023) calculations \cite{Kaneko2023} in the same framework for $Z=52$-56  corroborate these results, suggesting $N=76$ as the critical point of a prolate to oblate transition. 
 
These results are in qualitative agreement with earlier (2005) findings in the framework of the Z(4) CPS \cite{Bonatsos2005}, in which the parameter-independent predictions of this model have been compared to the data for $^{128-132}$Xe$_{74-78}$, indicating $^{130}$Xe$_{76}$ as the critical nucleus. 

These results are also compatible with the prediction of the proxy-SU(3) symmetry \cite{Bonatsos2017b} that the irreps corresponding to the valence neutrons in this shell become oblate from  $N=74$ onward (see Table III of \cite{Bonatsos2017b}), in accordance to the conclusion of Ref. \cite{Kaneko2023} that it is the intruder neutron partner orbitals ($1h_{11/2}$, $2f_{7/2}$) that are responsible for the prolate to oblate SPT at $N=76$. Within the proxy-SU(3) description, though, the total irreps representing the various nuclei become oblate only at $Z=72$ and beyond, since the irreps corresponding to the valence protons are prolate up to $Z=72$, thus preventing the total irreps to become oblate below $Z=72$ (see Table III of \cite{Bonatsos2017b}).  
 
In summary, there is growing evidence for a prolate to oblate transition at $N=76$. 

\subsection{The ${\bf Z\approx 44}$, ${\bf N\approx 64}$ region } \label{N64} 

Shell model calculations for the $N=66$ isotones \cite{Xu2002} suggest prolate shapes for $Z=34$-40 (i.e., for $^{100}$Se, $^{102}$Kr, $^{104}$Sr, and $^{106}$Zr), transitional behavior for $Z=42$, 44 (i.e., for $^{108}$Mo and $^{110}$Ru), and an oblate shape for $Z=46$ (i.e., for $^{112}$Pd). This finding is consistent with the proxy-SU(3) prediction that within the $Z=28$-50 shell, the first oblate irrep appears at 16 valence particles, i.e. at $Z=44$ (see Table I of \cite{Bonatsos2017b}). 

These findings suggest the existence of a prolate to oblate transition around $Z=44$ and/or $N=64$. 

 \subsection{The ${\bf Z\approx 34}$, ${\bf N\approx 34}$ region} \label{N34}
 
Shell model calculations for $N=Z$ nuclei in the $fpg$ shell \cite{Kaneko2004} provide prolate shapes for $N=Z=30$, 32 (i.e., for $^{60}$Zn$_{30}$ and $^{64}$Ge$_{32}$),  and oblate shapes at $N=Z=34$, 36 (i.e., for $^{68}$Se$_{34}$ and $^{72}$Kr$_{36}$). In other words, they suggest oblate shapes starting at 34 nucleons. This is in agreement with experimental results indicating an oblate shape for $^{68}$Se$_{34}$ \cite{Fischer2000}. 

In the proxy-SU(3) framework, nuclei in this region, in which protons and neutrons occupy the same shell, should be treated within the proxy-SU(4) symmetry, being under development by Kota \cite{Kota2023}.

Systematics of the energy difference $E(2_2^+)-E(4_1^+)$ suggest oblate shapes for $^{64}_{30}$Zn$_{34}$ and $^{72}_{32}$Ge$_{40}$ \cite{Castel1976}. Nilsson diagrams based on realistic Woods-Saxon potentials suggest oblate shapes for $^{64}_{28}$Ni$_{36}$ and $^{76}_{28}$Ni$_{48}$ \cite{Hamamoto2014}. 

In summary, there is some evidence for a prolate to oblate transition close to $N=34$ and/or $Z=34$. 

\subsection{Shape coexistence} 

Shape coexistence (SC) \cite{Heyde1983,Wood1992,Heyde2011,Heyde2016,Bonatsos2023c} in even-even nuclei refers to the situation in which the ground state band and another $K=0$ band lie close in energy but possess radically different structures, for example one of them being spherical and the other one deformed, or both of them being deformed, but one of them having a prolate shape and the other one exhibiting an oblate shape. A dual shell mechanism \cite{Martinou2021,Martinou2023} proposed within the proxy-SU(3) scheme \cite{Bonatsos2017a,Bonatsos2017b,Bonatsos2023a} suggests that SC can occur only within certain stripes of the nuclear chart, forming islands of SC, for the borders of which empirical rules have been recently suggested \cite{Bonatsos2023b}.  It is interesting to see where the lines along which a prolate to oblate transition is expected are lying in relation to the islands of SC, depicted in Fig. 1 of Ref. \cite{Bonatsos2023c}.

In subsecs. \ref{Pt}-\ref{Hg}, dealing with the region $Z=70$-80, the prolate to oblate transition is expected to occur at $N=116$, which lies outside the island of SC at these proton numbers. The same happens with the prolate to oblate transition at $N=76$, expected in the region $Z=50$-68, as described in subsec. \ref{Z5082}, as well as with the prolate to oblate transition around $Z=44$ described in subsec. \ref{N64}. 

In contrast, SC is observed close to the $Z=34$, $N=34$ region described in subsec. \ref{N34}, possibly suggesting further search for prolate to oblate transitions in additional medium-mass $N=Z$ nuclei. 

As already indicated in Fig. 1 of Ref. \cite{Bonatsos2023c}, SC is observed in several $N=28$ isotones. As we shall see in the next subsection, no prolate to oblate transitions have been observed in these isotones so far. The $N=28$ isotones call for further investigations in relation to a prolate to oblate transition, probably in analogy to the $N=64$ isotones, in which SC is known to occur but no clear evidence for a prolate to oblate transition exists yet, as discussed in subsec. \ref{N64}.  

From the microscopic point of view, SC is attributed \cite{Heyde1983,Wood1992,Heyde2011,Heyde2016,Bonatsos2023c} to particle-hole excitations across major shell or sub-shell closures \cite{Sorlin2008}, while recently particle-hole excitations across shell closures of the 3-dimensional isotropic harmonic oscillator have also been suggested \cite{Bonatsos2022a,Bonatsos2022b}, corroborated by covariant density functional theory calculations \cite{Bonatsos2022a,Bonatsos2022b}. However, in the case of the Pt isotopes it has been proved \cite{McCutchan2005} that a satisfactory description of their shape evolution, including the transition from prolate to oblate shapes,  can be obtained without using particle-hole excitations. This result might be indicating that particle-hole excitations are stronger near the $Z=82$ shell closure, i.e. in the Po, Pb, Hg ($Z=84$, 82, 80) series of isotopes, ``fading away'' as one moves away from the magic number $Z=82$ to the ``border case'' of Pt ($Z=78$).      

\subsection{The ${\bf N=28}$ isotones}
                                                                                                                                                                                                                                                                                                                                                                                                                                                                                                                                                                                                                                                                                                                                                                                                                                                                                                                                                                                                                                                                             The evolution \cite{Sorlin2008} of the magic number $N=28$ creates a special situation in some $N=28$ isotones. Early angular momentum projected calculations with the generator coordinate method have shown \cite{RodriguezGuzman2002} that the $N=28$ shell closure is preserved in $^{48}$Ca, but collapses in $^{40}$Mg, $^{42}$Si, $^{44}$S, and $^{46}$Ar, with shape coexistence predicted in the last two isotones. Subsequent calculations using the deformed Skyrme HFB \cite{Wang2014}, RMF+BCS \cite{Saxena2017}, and antisymmetrized molecular dynamics with the Gogny D1S density functional \cite{Kimura2013,Kimura2015,Suzuki2021,Suzuki2022} show that shape coexistence of prolate, oblate, and/or spherical shapes is expected in $^{40}$Mg, $^{42}$Si, and $^{44}$S. One may think that a prolate to oblate transition might occur in some of these $N=28$ isotones, but, to the best of our knowledge, no evidence exists so far in this direction.
 
\section{Experimental manifestations of O(6)} \label{O(6)}

In this section, existing experimental evidence for the O(6) symmetry is summarized, since O(6) has been suggested to be related to the prolate to oblate transition, reviewed in the previous section. Indeed, strong correlation between the two concepts is seen. 

The first experimental example provided for the O(6) symmetry in 1978, simultaneously with its discovery \cite{Arima1978,Arima1979}, is $^{196}$Pt$_{118}$ \cite{Cizewski1978}. A transition from O(6) to rotational behavior has been studied at the same time \cite{Casten1978b}, covering the nuclei $^{188-196}$Pt$_{110-118}$ and  
$^{186-194}$Os$_{110-118}$, with these nuclei considered as reasonable manifestations of O(6) \cite{Zamfir1998} (see subsec. III.B.1 of the review article \cite{Casten1988} for further discussion), while $^{198,200}$Pt$_{120,122}$ are considered to tend towards vibrational (U(5)) behavior and nuclei below $N=108$  are  considered as rotors close to the SU(3) behavior \cite{Zamfir1998}. Three-body interactions have been found \cite{Liao1994} to improve the agreement between the theoretical predictions of O(6) and the data for  $^{192-198}$Pt$_{114-120}$. In addition,  $^{196,198,202}$Hg$_{116,118,122}$ have been suggested as O(6) manifestations \cite{Morrison1981}. 

In summary, the $N=116$ point, identified in subsecs \ref{Pt}-\ref{Hg} as the critical point of a prolate to oblate transition, is found to lie within the region of O(6) DS in the Os-Pt region, in accordance with the theoretical considerations of \cite{Jolie2001,Warner2002,Jolie2002}. 
 
The second region of experimental manifestations of O(6) has been proposed in 1985, including $^{124-130}$Xe$_{70-76}$ and $^{128-134}$Ba$_{72-78}$ \cite{Casten1985a,Casten1985b,Casten1988}, with further features explored in $^{126}$Xe$_{72}$ \cite{Coquard2011}, $^{128}$Xe$_{74}$ \cite{Wiedenhover1997}, 
$^{128-132}$Ba$_{72-76}$ \cite{Zamfir1998}, and $^{134}$Ba$_{78}$ \cite{Molnar1988}.  Three-body interactions have been found \cite{Liao1994} to improve the agreement between the theoretical predictions of O(6) and the data for  $^{124-128}$Xe$_{70-74}$. Recent studies of entanglement entropy in $^{122-134}$Xe$_{68-80}$ \cite{Jafarizadeh2022} show increasing entanglement entropy (i.e, increasing O(6) character) from $^{128}$Xe$_{74}$ to $^{122}$Xe$_{68}$.  
$^{130}$Ce$_{72}$ has also been suggested as lying close to the O(6) symmetry, based on experimental lifetimes \cite{vonBrentano1988}. 
 
In summary, the $N=76$ point, identified in subsec. \ref{Z5082} as the critical point of a prolate to oblate transition, is found to lie within the region of O(6) DS in the Xe-Ba region, in accordance with the theoretical considerations of \cite{Jolie2001,Warner2002,Jolie2002}. 

\begin{figure}[htb] 

\includegraphics[width=75mm]{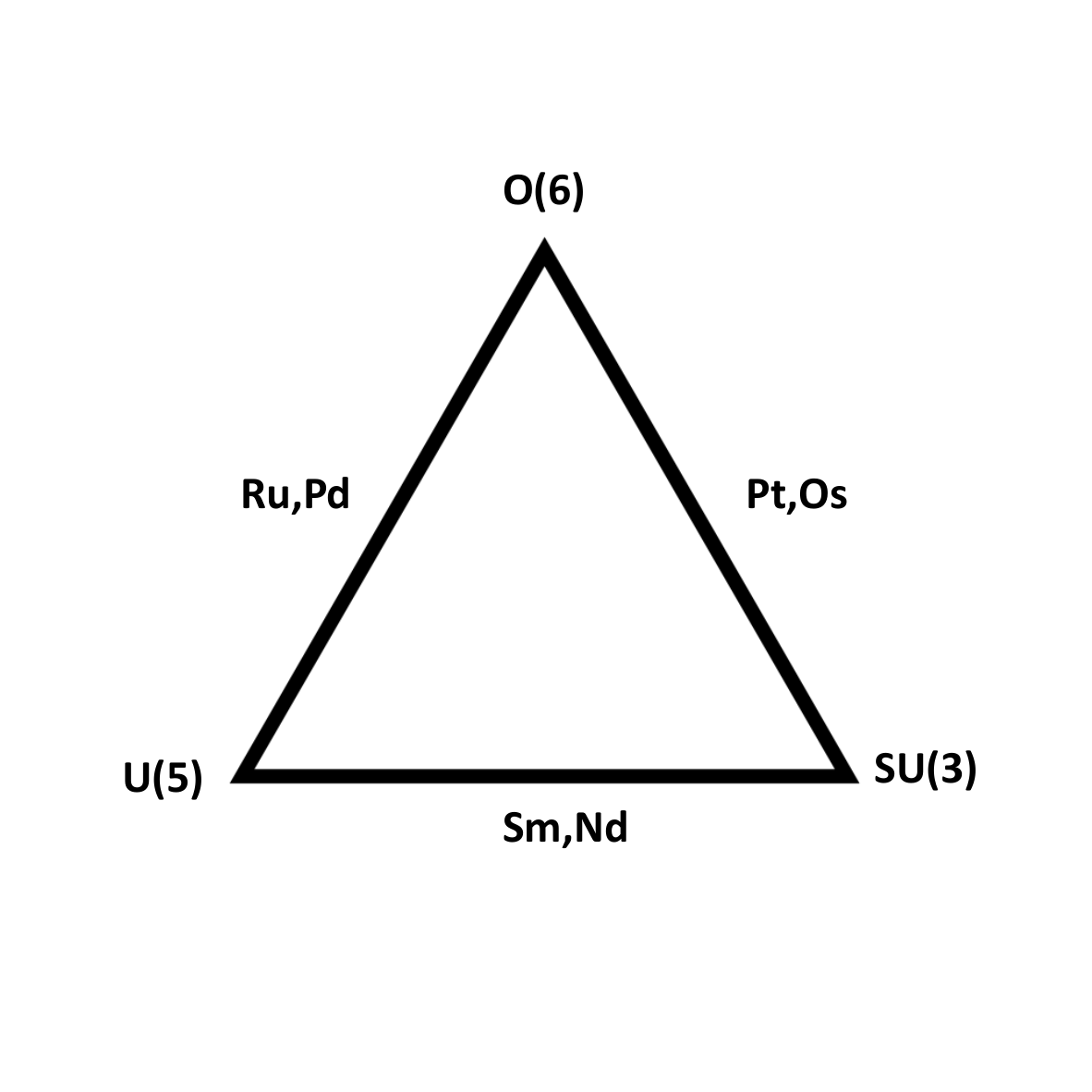}

\caption{Symmetry triangle of the IBM, depicting the nuclei representing experimental manifestations of the transitional regions between the three dynamical symmetries ((see Fig. 1 of \cite{Stachel1982}). See Sec. \ref{O(6)} for further discussion.
 }  
 \label{triangl3}

\end{figure}

The Ru ($Z=44$) and Pd ($Z=46$) isotopes have been studied in relation to the shape evolution from vibrational (U(5))  to $\gamma$-unstable (O(6)) shapes \cite{Stachel1982,Stachel1984,Casten1988,Soderstrom2013} (see Fig. \ref{triangl3}). It has been found that the Ru isotopes gradually approach the O(6) symmetry, with $^{108}_{44}$Ru$_{64}$ being the one closest to O(6) (see Fig. 9 of \cite{Soderstrom2013}).  The Pd isotopes also gradually approach the O(6) symmetry, with 
$^{110}_{46}$Pd$_{64}$ being the one closest to O(6) (see Fig. 5 of \cite{Stachel1982}). In other words, $N=64$ appears to be close to the critical point of a prolate to oblate SPT, the critical point exhibiting the O(6) symmetry. Recent studies of entanglement entropy in $^{102-110}$Pd$_{56-64}$ \cite{Jafarizadeh2022} show increasing entanglement entropy (i.e, increasing O(6) character) from $^{104}$Pd$_{58}$ to $^{110}$Pd$_{64}$.

In summary, the region around $Z=44$ and/or $N=64$ is found to lie within the region of O(6) DS in the Ru-Pd region, in accordance with the theoretical considerations of \cite{Jolie2001,Warner2002,Jolie2002} and the above mentioned (subsec. \ref{N64}) expectations for a prolate to oblate transition around $Z=44$ and/or $N=64$. 

IBM calculations including three-body interactions \cite{Liao1994} have suggested $^{72-76}_{32}$Ge$_{40-44}$ as O(6)-like nuclei. This suggestion is in agreement with the above mentioned (subsec. \ref{N34}) expectations for a prolate to oblate transition around $N=34$ and/or $Z=34$.

\section{Interplay between prolate to oblate transitions, O(6) and U(5)} \label{O6U5}
 
The findings of the previous section suggest that there is a strong correlation between the appearance of prolate to oblate SPTs and the occurrence of the  O(6) DS, in accordance with the expectations of Refs. \cite{Jolie2001,Warner2002,Jolie2002}. Furthermore, the nuclei best exhibiting both of these features, are located just below  closed proton and/or neutron shells, with oblate shapes appearing between these nuclei and close to the relevant magic numbers, in accordance to the predictions of the proxy-SU(3) \cite{Bonatsos2017b} and pseudo-SU(3) \cite{Bonatsos2020a} symmetries. In particular

a) The Os-Hg ($Z=76$-80) nuclei with $N=116$ lie below the $Z=82$  and $N=126$ shell closures, while oblate shapes are observed above $N=116$ until close to $N=126$. 

b) The Xe-Ba ($Z=54$-56) nuclei with $N=76$ lie below the $N=82$ shell closure,  while oblate shapes are observed above $N=76$ until close to $N=82$.

c) The Ru-Pd ($Z=44$-46) nuclei with $N=64$ lie below the $Z=50$ shell closure, with oblate nuclei observed above $Z=44$ until close to $Z=50$. 

It should be noticed that the appearance of oblate shapes below closed shells and prolate shapes above them is a universal effect, also appearing in alkali clusters 
\cite{deHeer1993,Brack1993,deHeer1987,Nesterenko1992}, which present magic numbers \cite{Knight1984,Martin1990,Martin1991,Bjornholm1990,Bjornholm1991,Pedersen1991,Brechignac1992,Brechignac1993}, and can be described by a Nilsson model without spin-orbit interactions \cite{Clemenger1985,Greiner1994} or by mean-field methods \cite{Brack1993,Nesterenko1992}. Relevant experimental observations can be seen in Refs. \cite{Borggreen1993,Pedersen1993,Pedersen1995,Haberland1999,Schmidt1999}.   
 
An important difference is that the transition from prolate to oblate shapes takes place through an O(6) critical point, the $\gamma$-unstable nature of which has been pointed out by several references in section \ref{po} (see, for example, \cite{Robledo2009,Alkhomashi2009}), while the transition from oblate shapes below magic numbers to prolate shapes above magic numbers takes place  through the region surrounding the magic numbers, which is expected to be spherical, thus characterized by the U(5) CPS.  In other words, the prolate to oblate transition takes place within a deformed, $\gamma$-unstable environment, while the oblate to prolate transition takes place through a spherical environment. This is corroborated by many sets of PESs derived through different approaches described in sec. \ref{meanfield}. 

The fact just mentioned also clarifies the difference between the U(5) and O(6) symmetries, which had been a point of discussion in the early days of the O(6) symmetry. Despite the fact that they share the common subalgebra O(5), which spans the whole line \cite{Leviatan1986} connecting U(5) to O(6) in the parameter space of the IBM (see Fig. \ref{triangl1}(a)), differences arise because of the different deformations. This was clarified in 1980 by the introduction of the classical limit of IBM \cite{Ginocchio1980a,Dieperink1980,Ginocchio1980b}, in which it became clear that the U(5) and O(6) symmetries  correspond to energy functionals,  the minima of which are 
$\gamma$-independent and correspond to $\beta=0$ and $\beta\neq 0$ respectively. Several years later this finding has been corroborated by the PES of mean field calculations mentioned in sec. \ref{meanfield}, which were not available at that time. In these PES it is clear that a deep valley connecting the prolate to oblate shapes is created away from $\beta=0$, corresponding to $\gamma$-unstable shapes which can accommodate O(6), while the prolate and oblate axes also meet at the $\beta=0$ point, representing the spherical U(5) case.   

\begin{figure}[htb] 

\includegraphics[width=75mm]{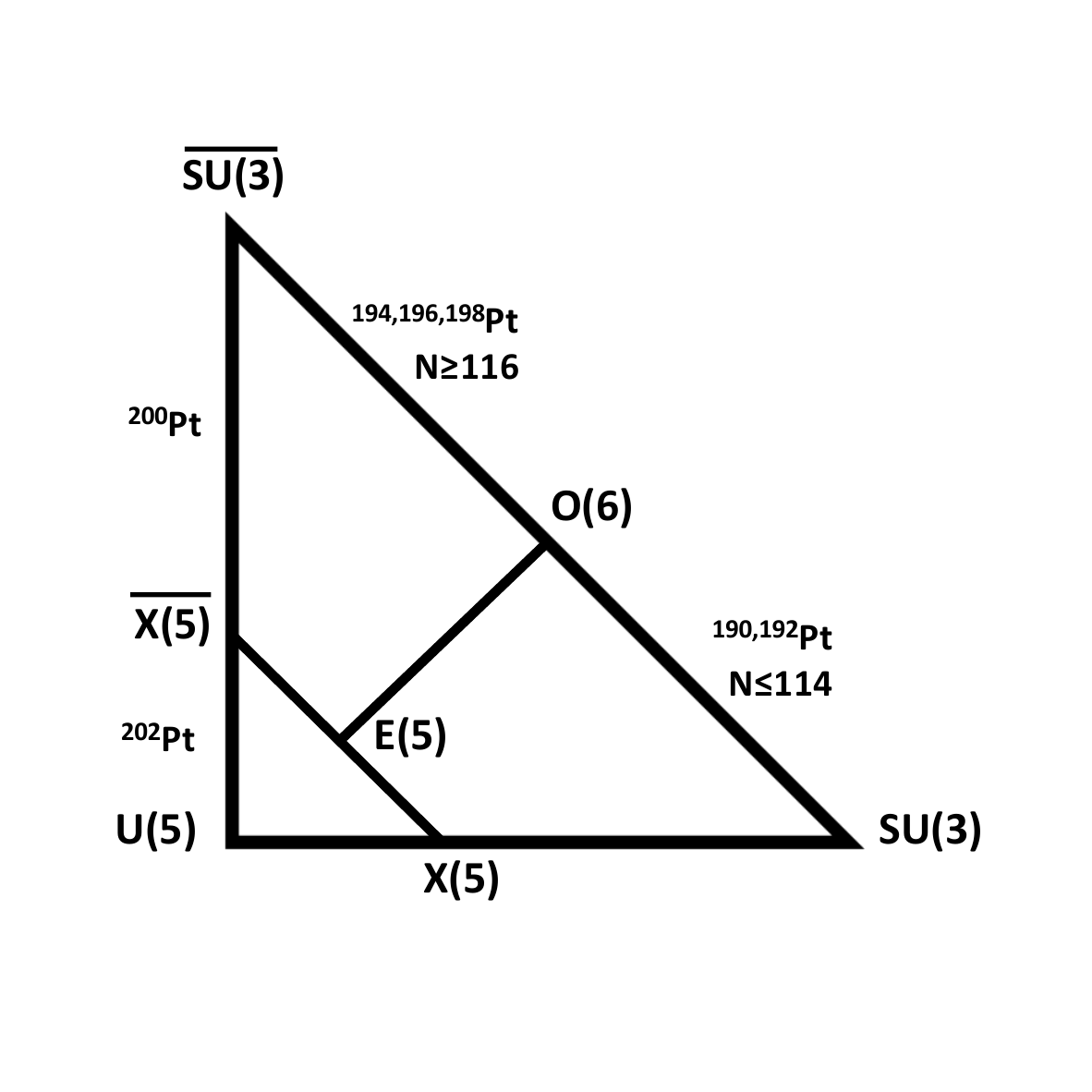}

\caption{Extended  symmetry triangle of the IBM, depicting the triple point (loosely labeled as E(5)) separating the spherical, prolate, and oblate phases, the critical point of the first order SPT between U(5) and SU(3) (loosely labeled as X(5)), the critical point of the first order SPT between U(5) and 
$\overline{\rm SU(3)}$ (loosely labeled as $\overline{X(5)}$), as well as O(6) as the critical point of the phase transition from prolate to oblate shapes. Nuclei representing experimental manifestations of the transitional regions are also shown. See Sec. \ref{O6U5} for further discussion.
 }  
 \label{around}

\end{figure}

Schematically in Fig. \ref{around} one can expect that the Pt isotopes $^{190,192}$Pt$_{112,114}$ lie close the SU(3)-O(6) leg, having prolate shapes, $^{194}$Pt$_{116}$ lies close to the O(6) critical point,  $^{196,198}$Pt$_{118,120}$ lie close to the O(6)-$\overline{\rm SU(3)}$ leg, having oblate shapes, while $^{200,202}$Pt$_{122,124}$ lie close to the $\overline{\rm SU(3)}$-U(5) leg. The light Pt isotopes with $N=90$-112  present a smooth evolution from near-spherical to quite deformed $\gamma$-unstable shapes, with maximum deformation achieved around midshell ($N=104$), with $^{182}$Pt$_{104}$ having the highest $R_{4/2}=E(4_1^+)/E(2_1^+)$ ratio (2.708) \cite{ENSDF}. Therefore they are expected to lie close to the U(5) to O(6) leg, but away from the critical point of the first order SPT, called X(5) in the collective model framework, since they do not present an abrupt transition from spherical to prolate deformed shapes. 

It should be remembered at this point that a mapping of the IBM parameters on the symmetry triangle of IBM has been introduced \cite{McCutchan2004}, by converting them into  polar coordinates. Using this mapping, the trajectories of the Gd-Hf \cite{McCutchan2004} and W-Pt isotopes \cite{McCutchan2005} within the symmetry triangle have been plotted, albeit only up to midshell ($N=104$). It might be instructive to pursue this task in  the upper half of the 82-126 neutron shell as well.  
 
The expectation of a transition from oblate to spherical shapes just  below closed shells has been recently pointed out by Kaneko \cite{Kaneko2023} using the quasi-SU(3) symmetry in the shell model, showing that $^{134}$Xe$_{80}$ is nearly spherical. A transition towards spherical shapes in the Pt isotopes as the $N=126$ shell closure is approached has also been pointed out by John \textit{et al.} \cite{John2017}.  
 
These findings indicate that a first order SPT from oblate deformed to spherical shapes takes place along the $\overline{\rm SU(3)}$-U(5) leg of the triangle of Fig. \ref{around}. This transition has been labeled by $\overline{\rm X(5)}$ in Ref. \cite{Warner2002}. No efforts have been made to identify this transition so far, since it seems that it takes place in a narrow region very close to the relevant neutron closed shell.  

Therefore the full picture around a shell closure indicates the existence of an oblate to spherical shape transition below the magic number, which can be called 
$\overline{\rm X(5)}$, and a transition from spherical to prolate shapes above the magic number, which corresponds to the X(5) CPS. This evolution is gradual, thus U(5) does not represent a critical point, in the way in which O(6) does in the prolate to oblate transition. 
 
\section{Two different paths from sphericity to deformation} \label{Qoper}  
 
 The differences between U(5) and O(6) mentioned in the previous section are reflected in the quadrupole operators of these two dynamical symmetries. The quadrupole operator of U(5), $(d^+\tilde d)^{(2)}$, contains only $d$ bosons, while the quadrupole operator of O(6), $(s^+\tilde d+ d^+ s)^{(2)}$, involves both $s$ and $d$ bosons, emphasizing the role of $s$ bosons in building up deformation. This role is emphasized by recent studies \cite{Jafarizadeh2022} of entanglement entropy in the IBM anf IBFM frameworks, showing that the entanglement between the $s$ and $d$ bosons along the U(5)-O(6) line is zero in the U(5) limit and maximum in the O(6) limit.  The two paths from sphericity to deformation are the topic of this section.
 
The U(5) limit of IBM \cite{Arima1976} corresponds to the Bohr model \cite{Bohr1952}. It contains only the 5 quadrupole bosons (the $d$ bosons), and is characterized by the U(5) DS, which is reduced as U(5)$\supset$O(5)$\supset$SO(3) \cite{Arima1976}. Thus it contains the seniority $\tau$ \cite{Rakavy1957} (characterizing the irreps of O(5)) and the angular momentum $L$ (characterizing the irreps of SO(3)) as good quantum numbers. In the classical limit, it corresponds to a spherical shape \cite{Iachello1987}. The relevant energy functional  has a single minimum at $\beta=0$ (see Sec. 3.4 of \cite{Iachello1987}). 

There are two ways to go away from the U(5) limit, by adding the monopole bosons (the $s$ bosons), and thus passing to the larger algebra U(6).  

The ``mild'' way to do it, is to preserve seniority as  a good quantum number. The relevant chain of subalgebras is  U(6)$\supset$O(6)$\supset$O(5)$\supset$SO(3) \cite{Arima1979}. In the classical limit it corresponds to an energy functional which is independent of $\gamma$ and has a single minimum at $\beta \neq 0$, which corresponds to a $\gamma$-unstable deformed shape \cite{Iachello1987}. For $N\to \infty$, the minimum occurs at $\beta=1$ (see Sec. 3.4 of \cite{Iachello1987}). 

The ``abrupt'' way to do it, is to break seniority.  The relevant chain of subalgebras is U(6)$\supset$SU(3)$\supset$SO(3) \cite{Arima1978b}.  In the classical limit it corresponds to an energy functional which has a minimum at $\gamma=0$ and $\beta\neq 0$, which corresponds to a prolate deformed shape \cite{Iachello1987}. For $N\to \infty$, the minimum occurs at $\beta=\sqrt{2}$ (see Sec. 3.4 of \cite{Iachello1987}). Thus within the SU(3) limit one can get larger deformations than in the O(6) limit. 

The ``mild'' path is known to correspond to a second order SPT, which in the Bohr framework has been called E(5) \cite{Iachello2000,Caprio2007}, it corresponds to the euclidean algebra in 5 dimensions and possesses the reduction chain E(5)$\supset$O(5)$\supset$SO(3). The separation of variables when solving the Schr\"{o}dinger equation for the E(5) SPT is exact, thus the obtained solution is exact \cite{Iachello2000,Caprio2007}. 

The ``abrupt'' path is known to correspond to a first order SPT, which in the Bohr framework has been called X(5) \cite{Iachello2001}. 
Its algebraic structure remains unknown. This is not surprising, since the separation of variables when solving the Schr\"{o}dinger equation for the X(5) SPT is 
approximate \cite{Iachello2001,Caprio2005},  thus the obtained solution is also approximate. In mathematical physics it is known that exact solutions of differential equations are possible when some appropriate symmetry is characterizing the Hamiltonian \cite{Dresner1999,Hydon2000}. It seems that no such symmetry exists in the X(5) case. Thus no appropriate algebra is expected to be ever found for X(5).

At this point it is instructive to consider the quadrupole operators. 

In U(5) \cite{Arima1976,Iachello1987} the quadrupole generator is $(d^\dagger\tilde d)^{(2)}$. This is the only possibility. The subalgebra O(5) is generated by  $(d^\dagger\tilde d)^{(3)}$ 
and  $(d^\dagger\tilde d)^{(1)}$, while SO(3) is generated by $(d^\dagger\tilde d)^{(1)}$ alone. Thus the quadrupole operator is involved only at the U(5) level and does not enter in its subalgebras \cite{Arima1976,Iachello1987}. 

In U(6), the quadrupole generators are $(d^\dagger\tilde d)^{(2)}$, $(d^\dagger\tilde s)^{(2)}$, and $(s^\dagger\tilde d)^{(2)}$ \cite{Iachello1987}. 

In the ``mild'' path, O(6) is generated by  $(d^\dagger\tilde s)^{(2)}+(s^\dagger\tilde d)^{(2)}$,  $(d^\dagger\tilde d)^{(3)}$ 
and  $(d^\dagger\tilde d)^{(1)}$ \cite{Arima1979,Iachello1987}. The next step is O(5), generated by $(d^\dagger\tilde d)^{(3)}$ 
and  $(d^\dagger\tilde d)^{(1)}$ alone. Thus the $s$ boson is not involved lower than the O(6) level, while seniority is preserved as a good quantum number below the O(6) level \cite{Arima1979,Iachello1987}. 

 In the ``abrupt'' path, SU(3) is generated by  $(d^\dagger\tilde s)^{(2)}+(s^\dagger\tilde d)^{(2)}\mp (\sqrt{7}/2)(d^\dagger\tilde d)^{(2)} $ and 
 $(d^\dagger\tilde d)^{(1)}$ \cite{Arima1978b,Iachello1987}. The next step is O(3), generated by  $(d^\dagger\tilde d)^{(1)}$ alone. Thus the $s$ boson is not involved lower than the SU(3) level, but senioriry is already gone at the SU(3) level \cite{Arima1978b,Iachello1987}. 

The difference between the two paths can be seen by looking at Fig. 3 of \cite{Bonatsos2004b}. The presence of O(5), and therefore of seniority as a good quantum number, has as a consequence that the spectrum consists of ``seniority trees'', consisting of an $L=0$ state with $\tau=0$, an $L=2$ state with $\tau=1$, a set of 
$L=4$,2 states with $\tau=2$, a set of $L=6$,4,3,0 states with $\tau=3$, and so on, as implied by Table I of \cite{Bonatsos2004b}. Thus in both the U(5) and O(6) limiting symmetries the spectrum consists of ``seniority trees''. This is not the case in SU(3). 

An important difference between the two paths is the following. Within U(5) and O(6), the quadrupole generator (which is $(d^\dagger\tilde d)^{(2)}$ in the case of U(5), while it is  $(d^\dagger\tilde s)^{(2)}+(s^\dagger\tilde d)^{(2)}$ in the case of O(6)), 
is \textsl{not} a generator of the underlying O(5) subalgebra, which is generated by $(d^\dagger\tilde d)^{(3)}$ 
and  $(d^\dagger\tilde d)^{(1)}$. As  a consequence, in both cases the quadrupole operator 
 breaks the O(5) symmetry, thus it can connect states  within bands of different seniority (different O(5) irreps) \cite{Arima1979,Iachello1987}. This is not the case in SU(3), since the quadrupole operator is indeed a generator of SU(3), thus it cannot break SU(3) and cannot connect states belonging to different SU(3) irreps \cite{Arima1978b,Iachello1987}. A more general quadrupole operator is needed in order to connect different SU(3) irreps. This more general operator should \textsl{not} be a generator of SU(3). One way to achieve this is to allow $\chi$ in Eq. (\ref{Qx}) to obtain values different from $\pm \sqrt{7}/2$, which correspond to the SU(3) symmetry \cite{Casten1988,Casten1990,Jolie2003}. 
 
\section{Two different paths from prolate  to oblate shapes} \label{Qoper2} 
 
 The last two sections clarify the two ways connecting prolate and  oblate shapes. 
 
 Considering a growing number of particles within a given major shell, in the beginning one has near-spherical nuclei characterized by the U(5) symmetry. Going through a first order SPT (called X(5) in the collective model framework) one reaches a region of prolate deformation, characterized by the SU(3) symmetry. Beyond the middle of the shell, the prolate to oblate SPT appears, which is an abrupt change from prolate to oblate shapes, predicted to have the O(6) symmetry. Advancing within the oblate region,  a little before the shell closure, a transition from oblate to spherical shapes occurs, for which the symbol $\overline{X(5)}$ can be used. Thus one starts with spherical shapes and ends up with spherical shapes again. In between, the prolate to oblate transition takes place, between the X(5) and $\overline{X(5)}$ transitions. The prolate to oblate transition is abrupt and is characterized by the O(6) CPS. Seniority is broken in the prolate SU(3) region before the O(6), as well as in the oblate $\overline{SU(3}$ region after O(6), but is reestablished at the critical point O(6). 
 
 Considering the transition from the region below a major shell closure to the region above it, on starts with an oblate deformed region, passes to a spherical region through $\overline{X(5)}$, crosses the magic number, still being in the spherical region, and then enters a prolate region by passing through the X(5) CPS. Around the magic numbers spherical shapes appear, characterized by the U(5) symmetry,   possessing the O(5) subalgebra. Thus senioriy is a good quantum number in the region between the $\overline{X(5)}$ and X(5) critical points. 
 
Therefore in both cases, when passing from prolate to oblate shapes, the seniority subalgebra O(5) is present. The difference is that the prolate to oblate transition takes place in a region with considerable deformation, characterized by the O(6) overall symmetry, while the oblate to prolate transition takes place in a spherical region, characterized by the U(5) overall symmetry. It should be remembered, however, that the spherical shape of doubly magic nuclei in the $pf$ shell has been recently questioned \cite{Poves2020} in the framework of configuration-interaction shell model calculations, by showing that large fluctuations are associated with the $\gamma$ variable in this case, thus rendering the "spherical" shape questionable. 
 
\section{Conclusions and outlook}

The main conclusions of the present review are summarized here.
 
The prolate to oblate shape transitions in rare earths at $N=116$ and $Z=76$, as well as at $Z\approx 44$, $N\approx 64$  are well established. They are predicted in a parameter-free way by the proxy-SU(3) and pseudo-SU(3) symmetries, provided that the highest weight irreducible representations of SU(3) are used in them. In addition, they are corroborated by non-relativistic and RMF calculations over series of isotopes, in which the parameters remain fixed throughout the nuclear chart, as well as by shell model calculations taking advantage of the quasi-SU(3) symmetry. 

The above mentioned regions, in which prolate to oblate shape transitions appear, coincide with regions in which experimental manifestations of the O(6) DS of the IBM have been observed, in agreement with the suggestion of the O(6) DS as the critical point of the transition from prolate to oblate shapes, proposed in the framework of the IBM. It is interesting that while seniority is not a good quantum number in prolate and oblate nuclei, it reappears as a good quantum number at the O(6) critical point between them. 

In addition, a gradual transition from oblate shapes appearing  below magic numbers to prolate shapes observed above magic numbers is seen in atomic nuclei, in analogy to alkali clusters. In atomic nuclei this transition goes through a U(5) region surrounding the magic numbers, leading to the conclusion that both prolate to oblate and oblate to prolate transitions take place through a $\gamma$-unstable region, which is O(6) and has non-zero deformation in the case of the prolate to oblate transition, while it is U(5) and has zero deformation for the oblate to prolate transition.  

In light nuclei, signs of a prolate to oblate transition appear in the region around $Z\approx 34$, $N\approx 34$, supported by shell model calculations taking advantage of the quasi-SU(3) symmetry, as well as by parameter-independent predictions of the proxy-SU(3) symmetry. The fact that protons and neutrons occupy the same major shell suggests that the proxy-SU(4) scheme \cite{Kota2023} should be applied in this case. 

Prolate and oblate shapes also appear in light nuclei \cite{Freer1995,Freer2021,Freer2022}, in which clustering becomes important. Since clustering in light nuclei has been recently reviewed in Ref. \cite{Freer2018}, this case has not been considered in the present review.  
 
\section*{Acknowledgements} 

Support by the Bulgarian National Science Fund (BNSF) under Contract No. KP-06-N48/1  is gratefully acknowledged.

\section*{Appendix: Acronyms} 

BCS Bardeen Cooper Schrieffer 

CDFT covariant density functional theory 

CPS critical point symmetry 

DS dynamical symmetry 

HF Hartree Fock 

HFB Hartree Fock Bogoliubov 

IBFM Interacting Boson Fermion Model 

IBM  Interacting Boson Model 

PEC potential energy curve 

PES potential energy surface 

PPQ pairing plus quadrupole 

RMF relativistic mean field 

SPT shape/phase transition 



\end{document}